\documentclass[11pt,a4paper,twoside]{article}
\usepackage{amssymb,amscd}
\usepackage{graphicx}
\usepackage{cite}
\usepackage{mathdots}
\usepackage{times}
\usepackage{pstricks,pst-node}
\newcommand{\be}{\begin{eqnarray}}
\newcommand{\ee}{\end{eqnarray}}
\newcommand{\bez}{\begin{eqnarray*}}
\newcommand{\eez}{\end{eqnarray*}}
\newcommand{\pa}{\partial}
\newcommand{\la}{\lambda}
\newcommand{\tr}{\mathrm{tr}}
\newcommand{\B}{\mathcal{B}}
\newcommand{\id}{\mathrm{id}}
\newcommand{\bt}{\mathbf{t}}
\newcommand{\bT}{\mathbf{T}}

\newcommand{\rmd}{\mathrm{d}}
\newcommand{\rmi}{\mathrm{i}}
\newcommand{\bd}{\bar{\mathrm{d}}}
\newcommand{\bD}{\bar{\mathrm{D}}}

\newcommand{\tv}{\tilde{\varphi}}

\title{\bf Dispersionless limit of the noncommutative potential KP hierarchy
 and solutions of the pseudodual chiral model in $2+1$ 
 dimensions\thanks{\copyright 2007 by A. Dimakis and F. M\"uller-Hoissen}
}

\author{Aristophanes Dimakis \\
 Department of Financial and Management Engineering, \\
 University of the Aegean, 31 Fostini Str., GR-82100 Chios, Greece \\
 dimakis@aegean.gr
          \and
 Folkert M\"uller-Hoissen \\
 Max-Planck-Institute for Dynamics and Self-Organization \\
 Bunsenstrasse 10, D-37073 G\"ottingen, Germany \\
 folkert.mueller-hoissen@ds.mpg.de }

\date{}

\begin{document}

\renewcommand{\theequation} {\arabic{section}.\arabic{equation}}

\newtheorem{theorem}{Theorem}
\newtheorem{lemma}{Lemma}
\newtheorem{proposition}{Proposition}
\newtheorem{definition}{Definition}
\newtheorem{corollary}{Corollary}
\newtheorem{propA}{Proposition A\hspace{-.1cm}}

\maketitle

\begin{abstract}
The usual dispersionless limit of the KP hierarchy does not work in the case where 
the dependent variable has values in a noncommutative (e.g. matrix) algebra. 
Passing over to the potential KP hierarchy, there is a corresponding scaling limit 
in the noncommutative case, which turns out to be the hierarchy of a ``pseudodual 
chiral model'' in $2+1$ dimensions (``pseudodual'' to a hierarchy extending 
Ward's (modified) integrable chiral model). 
Applying the scaling procedure to a method generating exact solutions of a 
matrix (potential) KP hierarchy from solutions of a matrix linear heat hierarchy, 
leads to a corresponding method that generates exact solutions of the matrix 
dispersionless potential KP hierarchy, i.e. the pseudodual chiral model 
hierarchy. We use this result to construct classes of exact solutions of 
the $su(m)$ pseudodual chiral model in $2+1$ dimensions, including various 
multiple lump configurations. 
\end{abstract}

\section{Introduction}
\label{section:intro}
Expressing the scalar KP hierarchy with dependent variable $u(t_1,t_2,\ldots)$
in terms of new evolution variables $T_n = \epsilon \, t_n$ with a parameter $\epsilon$,
the limit $\epsilon \to 0$ (keeping $T_n$ fixed) leads to the so-called 
dispersionless KP hierarchy 
(see \cite{Koda88,Koda+Gibb89,Kupe90qc,Kric92,Zakh94,Aoya+Koda94,Taka+Take95,Taka95,Stra95,Stra97,Chan+Tu00,Wieg+Zabr00,KMR01,DMT01,Duna+Tod02,Kono+Mart02,Kono+Mart02SIAM,MMM02,MMM02b,GMM03,Fera+Khus04,Mana+Sant06,Kono+Magr06}, for example).
The same limit does not work, however, for the KP hierarchy with dependent 
variable in a noncommutative (e.g. matrix) algebra.
In fact, different scaling limits of the matrix KP equation have already 
been explored in \cite{Zakh+Kuzn86}, where the multiscale expansion method 
has been used to relate different integrable systems.

In the present work we formulate a dispersionless limit of the 
``noncommutative'' potential KP (ncpKP) hierarchy with dependent variable $\phi$,  
where $u = \phi_{t_1}$. It turns out to be the hierarchy associated 
with a ``pseudodual chiral model'' (pdCM) in $2+1$ dimensions, a well-known 
reduction of the self-dual Yang-Mills equation \cite{Lezn87,Maso+Wood96}. 
Applying the scaling limit procedure to a method generating exact solutions 
of a matrix pKP hierarchy from solutions of a matrix linear heat hierarchy, 
then results in a method generating solutions of this pdCM hierarchy. 

In section~\ref{section:disp-pKP} we consider the dispersionless limit of the 
ncpKP equation. Section~\ref{section:disp-ncpKPhier} generalizes this limit to 
the whole ncpKP hierarchy, explores some of its properties, and in particular 
establishes a pseudoduality relation with a hierarchy that extends 
Ward's (modified) chiral model in $2+1$ dimensions \cite{Ward88,Ward88torsion,Ward89,Ward90,Ward94,Ward95}. 
The latter model has been studied extensively 
\cite{Lees89,Vill89,Sutc92,Sutc93,PSZ92,Ioan+Ward95,Ioan96,Ioan00,Ioan+Zakr98PLA,Ioan+Zakr98PLA2,Ioan+Zakr98JMP,Ioan+Mant05,Anand97,Anand98,ZMA05,Dai+Tern05,DTU06,Dai+Tern07,Duna+Mant05,Duna+Plan07}
(see also \cite{Ward99,Kote+Ward01,Ji+Zhou05} for the Ward model in 
(anti-) de Sitter space-time and 
\cite{Lech+Popo01b,Lech+Popo01c,Biel02,Wolf02,Ihl+Uhlm03,Chu+Lech05,Chu+Lech06,KLP06} for explorations of a Moyal-deformed version), in particular concerning 
its (multi-) lump solutions, which are two-dimensional soliton-like objects. 
In this respect, its pseudodual received comparatively little attention.  
The dependent variables of the two equations are related by a 
kind of hetero-B\"acklund transformation. Given a solution of one 
of the two equation, this becomes a first order system of partial differential equations, which determines a solution of the other equation. The 
necessary integration is typically difficult to carry out, however. 
Hence, although some properties of the pseudodual model can certainly 
be infered from corresponding knowledge of the Ward model, there 
is no \emph{explicit} translation of its solutions. In any case, 
in this work we present an independent approach to solutions of 
the pdCM and moreover to its hierarchy. 

In section~\ref{section:CHdisplimit} we derive the abovementioned method 
to generate exact solutions of the pdCM hierarchy from corresponding 
knowledge of the ncpKP hierarchy. The main result is independently 
verified in section~\ref{section:pdc-sols} and then applied to construct 
some classes of exact solutions. 
This section is actually formulated in such a way that it can be accessed 
almost without any knowledge of the previous sections. We concentrate 
on solutions of the $su(m)$ pdCM hierarchy and restrict concrete 
examples to the $su(2)$ case. 
Some conclusions are collected in section~\ref{section:conclusions}.

\section{The dispersionless limit of the noncommutative pKP equation}
\label{section:disp-pKP}
Let $\phi(\bt)$ with $\bt=(t_1,t_2,\ldots)$ be a function with values in some
matrix space\footnote{The entries will be taken as complex functions 
of $t_1,t_2,\ldots$, 
though large parts of this work also apply to the case where they are 
elements of any (possibly noncommutative) associative algebra, 
for which differentiability with respect to $t_1,t_2,\ldots$ can be defined.}
$\mathcal{A}$ which is endowed with a product $A \cdot B = A Q B$,
where $Q$ is a constant matrix, i.e. independent of $\bt$. 
We consider the following ncpKP equation,
\be
    4 \, \phi_{tx} - \phi_{xxxx} - 3 \, \phi_{yy}
  = 6 \, (\phi_x Q \phi_x)_x - 6 \, [\phi_x , \phi_y]_{Q} \, ,  \label{ncpKP}
\ee
where $x=t_1,y=t_2,t=t_3$ and
\be
   [A , B]_{Q} := A Q B - B Q A \; .
\ee
Let $\phi$ now also depend on a parameter $\epsilon$ in such a way that
\be
    \phi(\bt,\epsilon) = \epsilon^a \Phi(\bT) + \mathcal{O}(\epsilon^{a+1})  
                     \label{phi-expansion}
\ee
with some integer $a$. Furthermore, we assume that $Q$ has an expansion
\be
    Q = Q_{(0)} + \epsilon \, Q_{(1)} + \mathcal{O}(\epsilon^2) \; .
\ee
Rewriting the ncpKP equation in terms of the rescaled variables
$T_n = \epsilon \, t_n$, dividing the equation by the maximal power 
of $\epsilon$ common to all of its summands, 
and taking the limit $\epsilon \to 0$ while keeping $T_1,T_2,\ldots$ fixed, 
should result in an equation that still has linear as well as nonlinear 
terms (in $\Phi$). This fixes the value of $a$, but we have to distinguish 
the following two cases. 

If the algebra $(\mathcal{A},\cdot)$ is \emph{commutative} at $\epsilon=0$ 
with $Q_{(0)} \neq 0$, and hence the commutator 
$[\Phi_X , \Phi_Y]_{Q_{(0)}} = \Phi_X Q_{(0)} \Phi_Y - \Phi_Y Q_{(0)} \Phi_X$ 
vanishes, then our requirements lead to $a=-1$, and the scaling limit 
of the pKP equation, divided by $\epsilon$, is
\be
    4 \, \Phi_{TX} - 3 \, \Phi_{YY}
 =  6 \, (\Phi_X Q_{(0)} \Phi_X)_X - 6 \, [\Phi_X,\Phi_Y]_{Q_{(1)}} \, , 
\ee
where $X=T_1,Y=T_2,T=T_3$. 
If $\Phi$ is a scalar and $Q_{(0)}=1$, the last equation reduces to
\be
    4 \, \Phi_{TX} - 3 \, \Phi_{YY} = 6 \, (\Phi_X{}^2)_X \; . 
\ee
This is the potential form of the dispersionless limit of the (``commutative'') 
scalar KP equation, which is also known as the Khokhlov-Zabolotskaya equation 
(see \cite{Kric92} for instance). 

If the algebra $(\mathcal{A},\cdot)$ is \emph{noncommutative} at $\epsilon=0$, 
we have to set
\be
      a = 0 \, ,
\ee
and this choice will be made throughout this work. 
Then we obtain the following \emph{dispersionless limit of the ncpKP equation} (\ref{ncpKP}), 
\be
    4 \, \Phi_{TX} - 3 \, \Phi_{YY} = - 6 \, [\Phi_X , \Phi_Y]_{Q_{(0)}} \; .
            \label{disp-ncpKP}
\ee
Up to the modified matrix product and rescalings of the coordinates, this is a 
well-known reduction of the self-dual {Y}ang-{M}ills equation 
(see \cite{Lezn87,Lezn+Mukh87,Lezn+Save89,Papa89,Papa91recur,Maso+Wood96,Foka+Ioan01}). 
With the further dimensional reduction $\Phi_X = \Phi_T$, it becomes the 
\emph{pseudodual chiral model} 
\cite{Frad+Tsey85,Curt+Zach94,Curt+Zach95para} (see also 
\cite{Zakh+Mikh78rel,Napp80,Frid+Jevi84,Lezn+Mukh87}). Accordingly, 
we may call (\ref{disp-ncpKP}) a \emph{pseudodual chiral model in $2+1$ dimensions}, 
in the following abbreviated to \emph{pdCM}. 
In fact, as explained in section~\ref{subsec:dncpKP_Ward}, it is ``pseudodual'' 
to an integrable (modified) chiral model in $2+1$ dimensions.

\section{The dispersionless limit of the ncpKP hierarchy}
\label{section:disp-ncpKPhier}
A functional representation of the ncpKP hierarchy is given by \cite{DMH07Burgers}
\be
    (\phi - \phi_{-[\la]})(\la^{-1} - Q \phi) - \phi_{t_1} 
  = \theta - \theta_{-[\la]} \, ,
\ee
where $\theta$ is an arbitrary $\mathcal{A}$-valued function, and
$(\phi_{-[\la]})(\bt) := \phi(\bt - [\la])$ is a Miwa shift with
$[\la]=(\la,\la^2 /2,\la^3/3,\ldots)$, $\la$ an indeterminate.
Eliminating $\theta$ from this equation, we get the following functional form
of the ncpKP hierarchy,
\be
 && \Big((\phi-\phi_{-[\la]})(\la^{-1} - Q \phi) - \phi_{t_1}\Big)
     - \Big((\phi-\phi_{-[\la]})(\la^{-1} - Q \phi) - \phi_{t_1}\Big)_{-[\mu]} 
             \nonumber \\
 &=& \Big((\phi-\phi_{-[\mu]})(\mu^{-1} - Q \phi) - \phi_{t_1}\Big)
     - \Big((\phi-\phi_{-[\mu]})(\mu^{-1} - Q \phi) 
     - \phi_{t_1}\Big)_{-[\la]} \, , 
\ee
where $\mu$ is another indeterminate.

If $\mathbf{p}_n$, $n=1,2,\ldots$, denote the elementary Schur polynomials, then
\be
    \mathbf{p}_n(-\tilde{\pa}) 
  = - \frac{\epsilon}{n} \, \pa_{T_n} + \mathcal{O}(\epsilon^2) \, ,
\ee
where $\tilde{\pa} = (\pa_{t_1},\pa_{t_2}/2,\pa_{t_3}/3,\ldots)$, and hence 
\be
    \phi - \phi_{-[\la]} = \epsilon \, D(\la) \Phi + \mathcal{O}(\epsilon^2) \, ,
\ee
where
\be
    D(\la) := \sum_{n \geq 1} {\la^n \over n} \, \pa_{T_n} \; .
\ee
In accordance with (\ref{phi-expansion}), where now $a=0$, we shall assume
\be
    \theta(\bt,\epsilon) = \Theta(\bT) + \mathcal{O}(\epsilon) \; .
\ee
Then we obtain
\be
    D(\la)(\Phi) \, (\la^{-1} - Q_{(0)} \Phi) - \Phi_{T_1} = D(\la)(\Theta) \; .
\ee
Expanding this in powers of $\la$, we find
\be
    {1\over n+1} \Phi_{T_{n+1}} - {1 \over n} \Phi_{T_n} Q_{(0)} \Phi
  = {1\over n} \Theta_{T_n}    \qquad n=1,2 \ldots \; .
        \label{Phi-Theta-hier}
\ee
Elimination of $\Theta$ results in the hierarchy equations
\be
    {n \over (n+1)} \Phi_{T_{n+1},T_m} - {m \over (m+1)} \Phi_{T_{m+1},T_n}
  = \Phi_{T_n} Q_{(0)} \Phi_{T_m} - \Phi_{T_m} Q_{(0)} \Phi_{T_n} \; .
\ee
Introducing 
\be
     x_n := n \, T_n \qquad \quad n=1,2,\ldots \, ,  \label{x_n-T_n}
\ee
this becomes
\be
    \Phi_{x_{n+1},x_m} - \Phi_{x_{m+1},x_n}
  =  [ \Phi_{x_n} , \Phi_{x_m} ]_{Q_{(0)}}  \qquad\quad m,n=1,2,\ldots \; . 
      \label{disp-ncpKPhier}
\ee
For $m=1, n=2$, we recover (\ref{disp-ncpKP}). 

Expressing $Q_{(0)}$ as 
\be
           Q_{(0)} = V U^\dagger   \label{Q_0VU}
\ee 
with matrices $U,V$ and the adjoint (complex conjugate and transpose) 
$U^\dagger$ of $U$, then 
\be
    \varphi := U^\dagger \Phi V      \label{varphi-UV}
\ee
(which includes the cases $\varphi = Q_{(0)} \Phi $ and 
$\varphi = \Phi Q_{(0)}$) solves
\be
    \varphi_{x_{n+1},x_m} - \varphi_{x_{m+1},x_n} =
    [\varphi_{x_n},  \varphi_{x_m}]  \qquad\quad m,n=1,2,\ldots 
    \, ,       \label{varphi_hier}
\ee
if $\Phi$ solves (\ref{disp-ncpKPhier}).
The power of this observation lies in the fact that any solution of (\ref{disp-ncpKPhier}) 
in some $M \times N$ matrix algebra, where $Q_{(0)} = V U^\dagger$ with an 
$M \times m$ matrix $U$ and an $N \times m$ matrix $V$, determines in 
this way a solution of (\ref{varphi_hier}) in the $m \times m$ matrix algebra. 
For example, if we are looking for solutions of (\ref{varphi_hier}) 
in the algebra of $2 \times 2$ matrices, we may first look for solutions 
of (\ref{disp-ncpKPhier}) with \emph{any} $M,N \geq 2$ and 
$Q_{(0)} = V U^\dagger$ with $M \times 2$ and $N \times 2$ matrices $U$ 
and $V$. In this way (simple) solutions of (\ref{disp-ncpKPhier}) in arbitrarily 
large matrix algebras lead to (complicated) solutions of (\ref{varphi_hier}) in 
the algebra of $2 \times 2$ matrices. 
In particular, this explains the significance of $Q_{(0)}$ in our previous formulae.  
In section~\ref{section:pdc-sols} we will substantiate this method. 
The hierarchy (\ref{varphi_hier}) is consistent with restricting $\varphi$ to 
take values in any Lie algebra, e.g. $sl(N,\mathbb{R})$, $sl(N,\mathbb{C})$, $u(N)$ 
or $su(N)$. If $\varphi$ solves (\ref{varphi_hier}), then also $\varphi + \varphi_0$, 
where $\varphi_0$ is a constant in the respective Lie algebra. 

As a consequence of their origin, the hierarchies (\ref{disp-ncpKPhier}) 
and (\ref{varphi_hier}) are invariant under the scaling transformation 
$x_n \mapsto \lambda \, x_n$, $n=1,2,\ldots$, with any constant $\lambda \neq 0$.
\vskip.1cm

\noindent
\textbf{Remark.} If $g_1,g_2$ are any two constant invertible matrices with 
size such that $g_1 \Phi g_2$ is defined, then 
\be
    \Phi \mapsto g_1 \Phi g_2 \, , \qquad 
    Q_{(0)} \mapsto g_2^{-1} Q_{(0)} g_1^{-1}       \label{Phi-gaugetransf}
\ee
leaves (\ref{disp-ncpKPhier}) invariant. If $Q_{(0)}$ is given by 
(\ref{Q_0VU}), the latter transformation results from 
\be
      V \mapsto g_2^{-1} V \, , \qquad U \mapsto (g_1^\dagger)^{-1} U \, , 
\ee
and $\varphi$ is invariant. More generally, the transformation 
$V \mapsto g_2^{-1} V \sigma$, $U \mapsto (g_1^\dagger)^{-1} U (\sigma^\dagger)^{-1}$, with a constant $m \times m$ matrix $\sigma$,  
leads to $\varphi \mapsto \sigma^{-1} \varphi \sigma$. This leaves the 
hierarchy equations (\ref{varphi_hier}) invariant. 
\hfill $\square$

\subsection{Some properties of the first dispersionless hierarchy equation}
\label{subsec:dncpKP_prop}
A Lagrangian for the first equation ($m=1,n=2$) 
\be
   \varphi_{x_1,x_3} - \varphi_{x_2,x_2} = - [\varphi_{x_1},  \varphi_{x_2}]
        \label{1st-eq}
\ee
of the hierarchy (\ref{varphi_hier}) is 
\be
     \mathcal{L} 
  = -\tr \Big( \varphi_{x_1}\varphi_{x_3} 
     - \varphi_{x_2}{}^2 -{2 \over 3}\varphi [\varphi_{x_1},\varphi_{x_2}] \Big) 
\ee
(see also \cite{Lezn+Mukh87,Lezn+Save89}). After passage to the new coordinates 
$x,y,t$ given by
\be
    x_1 = \frac{1}{2} (t-x) \, , \qquad
    x_2 = y \, , \qquad
    x_3 = \frac{1}{2} (t+x) \, ,    \label{cord_xyt}
\ee
equation (\ref{1st-eq}) becomes
\be
 \varphi_{tt} - \varphi_{xx} - \varphi_{yy} + [ \varphi_t - \varphi_x , \varphi_y ] 
   = 0 \, ,   \label{1st-eq-xyt}
\ee
and the Lagrangian takes the form
\be
  \mathcal{L} = - \frac{1}{2} \tr \Big( \varphi_t{}^2 - \varphi_x{}^2 - \varphi_y{}^2
                  - {2 \over 3} \varphi [\varphi_t-\varphi_x,\varphi_y] \Big) 
                       \nonumber \\
   = -\frac{1}{2} \tr\Big( \eta^{\mu \nu} \pa_\mu\varphi \, \pa_\nu\varphi 
      + \frac{2}{3} \varphi \, v_\rho \epsilon^{\rho\mu\nu} 
      \pa_\mu\varphi \, \pa_\nu\varphi \Big) \, , 
\ee
where we introduced the components $\eta^{\mu \nu}$ (with respect to the 
coordinates $(x^\mu)=(t,x,y)$) of the Minkowski metric in $2+1$ dimensions, 
the totally antisymmetric Levi-Civita pseudo-tensor with $\epsilon^{012} = 1$, 
and a constant covector $v_\rho$ with components $(1,1,0)$. 
As a consequence of the translational invariance of the Lagrangian, 
the energy-momentum tensor 
\be
    T^\mu{}_\nu = \tr\Big( \frac{\pa \mathcal{L}}{\pa(\pa_\mu\varphi)} 
       \, \pa_\nu\varphi - \delta^\mu_\nu \, \mathcal{L} \Big)
\ee
provides us with the conserved densities
\be
  &&  T^0{}_0 = - \frac{1}{2} \tr\Big( \varphi_t^2+\varphi_x^2+\varphi_y^2 
      - \frac{2}{3} \varphi [\varphi_x,\varphi_y] \Big) \, , 
            \cr
  &&  T^0{}_1 = - \tr\Big( \varphi_t \varphi_x 
     - \frac{1}{3} \varphi [\varphi_x,\varphi_y] \Big) \, , \qquad
    T^0{}_2 = - \tr(\varphi_t \varphi_y) \; .
\ee
Then also
\be
    \mathcal{E} 
  = T^0{}_0 -T^0{}_1
  = - {1 \over 2} \tr[(\varphi_t - \varphi_x)^2 + \varphi_y{}^2] 
          \label{energy}
\ee
is a conserved density. For any non-zero \emph{anti-Hermitian} matrix, 
the trace of the square of the matrix is real and negative. 
Hence $\mathcal{E}$ provides us with a \emph{non-negative}  
``energy'' density in the case where $\varphi$ takes values in the Lie algebra 
$u(m)$ of the unitary group.

For any infinitesimal symmetry  
$\delta \varphi = \frac{\pa\varphi}{\pa \alpha} \delta \alpha$ 
(with a parameter $\alpha$) of the Lagrangian, there is a conserved current 
\be
     J^\mu := \tr\Big( \frac{\pa\mathcal{L}}{\pa(\pa_\mu\varphi)} 
              \frac{\pa \varphi}{\pa\alpha} \Big) \, , 
\ee
i.e., $\pa_\mu J^\mu =0$. 
A symmetry of the above Lagrangian is given by 
$\delta \varphi = [C,\varphi] \alpha$
with any constant (anti-Hermitian) matrix $C$. Hence 
\be
  J^0_C = - \tr\left( \Big( [\varphi,\varphi_t] 
   + \frac{1}{3} ( \varphi^2 \varphi_y 
   - 2 \varphi \varphi_y \varphi + \varphi_y \varphi^2) \Big) \, C \right) 
\ee
is a conserved density.

\subsection{Relation with Ward's chiral model in $2+1$ dimensions}
\label{subsec:dncpKP_Ward}
The hierarchy (\ref{varphi_hier}) is related to the hierarchy of an integrable 
(modified) chiral model in $2+1$ dimensions. 
First we note that (\ref{varphi_hier}) is the integrability condition of the 
linear system
\be
    J_{x_{n+1}} = - J \, \varphi_{x_n}    \qquad \quad n=1,2,\ldots
                 \label{J-varphi-eq}
\ee
with some invertible $J$. Rewriting this as 
\be
    \varphi_{x_n} = - J^{-1} J_{x_{n+1}}  \qquad \quad n=1,2,\ldots \, , 
        \label{varphi-J-eq}
\ee
we find that (\ref{varphi_hier}) is automatically satisfied and the integrability 
conditions now take the form 
\be
    (J^{-1} J_{x_{n+1}})_{x_m} - (J^{-1} J_{x_{m+1}})_{x_n} = 0 
    \qquad \quad  m,n = 1,2, \ldots \, \; .
           \label{cmodel_hier}
\ee
In conclusion, solutions $J$ of (\ref{cmodel_hier}) are in correspondence with 
solutions $\varphi$ of (\ref{varphi_hier}) via (\ref{varphi-J-eq}). 
This correspondence is of a nonlocal nature. In particular, given a 
solution $\varphi$ of (\ref{varphi_hier}), (\ref{varphi-J-eq}) does 
not directly determine $J^{-1} J_{x_1}$. 
We first have to solve (\ref{J-varphi-eq}) for $J$ in order to be able to 
calculate this expression.

(\ref{cmodel_hier}) is immediately recognized as the dispersionless limit of the 
noncommutative \emph{modified} KP hierarchy (see equation (4.12) in \cite{DMH06func}).

 For $m=1$ and $n=2$, (\ref{cmodel_hier}) reads 
\be
    (J^{-1} J_{x_3})_{x_1} - (J^{-1} J_{x_2})_{x_2} = 0 \; .  \label{Ward-CM-eq}
\ee
This equation apparently first appeared in \cite{Pohl80,Mana+Zakh81}. 
It is a reduction of the self-dual Yang-Mills equation 
(see \cite{Pohl80,Lezn+Save89,Maso+Wood96}, for example).
In terms of the coordinates $x,y,t$ given by (\ref{cord_xyt}), it takes the form
\be
  (J^{-1} J_t)_t - (J^{-1} J_x)_x - (J^{-1} J_y)_y + [J^{-1} J_x , J^{-1} J_t] = 0 \, ,
\ee
or in tensor notation (using the summation convention)
\be
    (\eta^{\mu \nu} + \epsilon^{\mu \nu}) \pa_\mu ( J^{-1} \pa_\nu J) = 0 \, ,
          \label{Ward-chiralmodel}
\ee
where $\mu,\nu = 0,1,2$, $(\eta^{\mu \nu}) = \mathrm{diag}(1,-1,-1)$, and $\epsilon^{\mu \nu}$
is antisymmetric with $\epsilon^{01} = - \epsilon^{10} = 1$ and zero 
otherwise. We note that the bivector $\epsilon^{\mu\nu}$ breaks 
Lorentz invariance in $2+1$ dimensions. Using the Lorentz invariant 
Levi-Civita pseudo-tensor and the constant unit covector $v_\alpha$ 
with components $(0,0,1)$, it can be expressed as 
$\epsilon^{\mu\nu} = v_\alpha \epsilon^{\alpha \mu \nu}$. Another integrable 
equation is obtained if we choose $v_\alpha$ to be timelike 
\cite{Mana+Zakh81,Ward88,Ioan+Ward95}. (\ref{Ward-chiralmodel}) 
is Ward's $(2+1)$-dimensional generalization of the chiral (or sigma) model \cite{Ward88,Ward88torsion,Ward89,Ward90,Ward94,Ward95}, see also \cite{Lees89,Vill89,Sutc92,Sutc93,PSZ92,Ioan+Ward95,Ioan96,Ioan00,Ioan+Zakr98PLA,Ioan+Zakr98PLA2,Ioan+Zakr98JMP,Foka+Ioan01,Ioan+Mant05,Anand97,Anand98,ZMA05,Dai+Tern05,Dai+Tern07,Duna+Mant05,Duna+Plan07}.
$J$ can be consistently restricted to any Lie group, e.g. $SL(N,\mathbb{R})$, 
$SL(N,\mathbb{C})$, $U(N)$ or $SU(N)$. 
\vspace{.1cm}

\noindent
\textbf{Remark.} According to (\ref{varphi-J-eq}) we have
$J^{-1} J_ y  = \varphi_x - \varphi_t$ and $J^{-1} J_t + J^{-1} J_x = -\varphi_y$, 
in terms of the variables $x,y,t$ given by (\ref{cord_xyt}). Hence 
\be
   \mathcal{E} = \mathcal{E}_{\mathrm{Ward}} 
       - \tr\left( J^{-1} J_t \, J^{-1} J_x \right) \,,   \label{E-E_Ward}
\ee
where 
\be
   \mathcal{E}_{\mathrm{Ward}} 
 = - \frac{1}{2} \tr\left( (J^{-1}J_t)^2 + (J^{-1} J_x)^2 + (J^{-1}J_y)^2 \right) 
\ee
is the energy density of Ward's chiral model. The difference between 
$\mathcal{E}_{\mathrm{Ward}}$ and $\mathcal{E}$ is \emph{not} a 
\emph{local} expression in terms of $\varphi$. 
The appendix attempts to further clarify the relation between 
Ward's chiral model and the pdCM hierarchy (and yet another version of it). 
\hfill $\square$

\subsection{An associated bidifferential calculus}
\label{subsec:bidiff}
On the algebra $\mathcal{A}$ of $m \times m$ matrices with entries depending 
smoothly on $x_1,x_2,\ldots$, 
we introduce two linear maps $\rmd, \bd$ by\footnote{We note that 
$\bd = \mathcal{R} \circ \rmd$ where $\mathcal{R}$ is the linear 
left $\mathcal{A}$-module map determined by $\mathcal{R}(\rmd x_n)=\rmd x_{n-1}$ 
for $n>1$, and $\mathcal{R}(\rmd x_1)=0$. This makes contact 
with Fr\"olicher-Nijenhuis theory \cite{Froe+Nije56}, see also \cite{CST00}.} 
\be
    \rmd \psi = \sum_{n\geq 1} \psi_{x_n} \, \rmd x_n \, , \qquad
    \bd \psi = \sum_{n\geq 1} \psi_{x_{n+1}} \, \rmd x_n \; . 
\ee
By use of the graded Leibniz rule they extend to a (bi-) differential 
graded algebra and satisfy 
\be
    \rmd^2 = \bd^2 = \rmd \bd + \bd \rmd = 0 \, ,  \label{bicomplex}
\ee
and hence we have a \emph{bidifferential calculus}. Dressing $\bd$ by setting
\be
    \bD \psi = \bd \psi - A \, \psi \, ,
\ee
with a 1-form $A = \sum_{n \geq 1} A_{n+1} \rmd x_n$, we find that 
$\rmd,\bD$ yields again a bidifferential calculus ($\bD^2 = \rmd \bD + \bD \rmd = 0$), 
iff
\be
    \rmd A = 0 \, , \qquad \bd A = A \wedge A   \label{A-eqs-bidiff}
\ee
(see also \cite{DMH00a}).
These equations cover Ward's chiral model hierarchy as well as its pseudodual, 
which is the dispersionless ncpKP hierarchy. Indeed, solving the first equation 
by setting
\be
    A = \rmd \varphi \, ,    \label{A=dvarphi}
\ee
the second reproduces the pdCM hierarchy
\be
    \bd \rmd\varphi = \rmd\varphi \wedge \rmd\varphi \; .
\ee
Alternatively, solving the second of equations (\ref{A-eqs-bidiff}) by setting
\be
    A = - J^{-1} \bd J \, ,   \label{A=-J^-1bdJ}
\ee
we recover the hierarchy
\be
    \rmd(J^{-1} \bd J) = 0    \label{Ward-hier}
\ee
associated with Ward's chiral model. 
The relation between both hierarchies is given by 
\be
    J^{-1} \bd J = - \rmd \varphi   \label{varphi-J-eq2}
\ee
(which is (\ref{varphi-J-eq})). 
This may be regarded as a ``Miura transformation''. 
The linear system associated with the bidifferential calculus is 
\be
    \bD \psi - \la \, \rmd \psi = 0 \, ,
\ee
with a parameter $\la$. Taking components of the differential forms, 
this reads
\be
    (\pa_{x_{n+1}} - A_{n+1} - \la \pa_{x_n}) \psi = 0  \qquad\quad n = 1,2,\ldots \;.
\ee
The integrability conditions now have the form
\be
  [\pa_{x_{n+1}} - A_{n+1} - \la \pa_{x_n} , \pa_{x_{m+1}} 
     - A_{m+1} - \la \pa_{x_m}] = 0 \; .
\ee
Its multicomponent version (and with $m,n \in \mathbb{Z}$) appeared in 
\cite{Taka90} (see (2.1), (2.2), and also the references therein).

Nonlocal conserved currents are obtained in the following way \cite{DMH00a}. 
Let $\rmd \chi_0 =0$. As a consequence of the bidifferential calculus 
structure, there are $\chi_n$, $n=1,2,\ldots$, such that
\be
   j_{n+1} := \bD \chi_n = -\rmd \chi_{n+1}  \qquad \quad n=0,1,\ldots
\ee
iteratively determines $\chi_n$, $n=1,2,\ldots$. 
For example, starting with $\chi_0 = \mathcal{I}$ (the unit matrix), 
we get $j_1 = \bD \mathcal{I} = -\rmd\varphi$ (using (\ref{A=dvarphi})), 
hence $\chi_1 =\varphi + a$ with $\rmd a = 0$, and thus also $\bd a = 0$. 
In the second step we have 
$j_2 = \bD(\varphi+a) = \bd \varphi - \rmd\varphi \, (\varphi+a)$, 
and the construction of the next current requires the integration
of $\rmd\chi_2 = \rmd\varphi \, (\varphi+a) - \bd \varphi$. The constant $a$ 
actually turns out to be redundant and should be set to zero.

A B\"acklund transformation is obtained from
\be
    (\rmd-\la^{-1} \bD)(\mathcal{I}+\la^{-1} \B) 
  = (\mathcal{I}+\la^{-1} \B)(\rmd -\la^{-1}\bD') \, , \quad 
    \bD' = \bd - A' \, , 
\ee 
with an operator $\B$ (see \cite{DMH01bt}). Expanding in powers of 
$\la^{-1}$, we find 
\be
    [\rmd ,\B] = \bD - \bD' \, , \qquad \bD \B = \B \bD' \; . 
\ee
Assuming $\B(\psi) = B \psi$ with a matrix $B$, this means
\be
    \rmd(B) = A' -A \, , \qquad \bd(B) = A B - B A' \; .
        \label{BT}
\ee
Using (\ref{A=dvarphi}) and solving the first of these equations by 
setting $B = \varphi' -\varphi -a$ with $\rmd a=0$, we obtain from the second
\be
    \bd(\varphi' - \varphi) = \rmd\varphi \, (\varphi' -\varphi -a) 
           - (\varphi' -\varphi -a) \, \rmd \varphi' \, , 
\ee
a B\"acklund transformation of the pdCM hierarchy. 
Alternatively, using (\ref{A=-J^-1bdJ}) and solving the second of 
equations (\ref{BT}) by setting
$B = -J^{-1}\mathcal{K} J'$ with $\bd \mathcal{K}=0$, the first becomes 
\be
    J'{}^{-1} \bd J' - J^{-1} \bd J = \rmd(J^{-1} \mathcal{K} J') \, , 
\ee
a B\"acklund transformation of the (modified) chiral model hierarchy (see 
also \cite{OPSC80} for the case of the chiral model on a two-dimensional 
space-time).

If $\B_{ij}$ leads from an $i$th to a $j$th solution, 
a permutability relation is given by
\be
    B_{12} + B_{24} = B_{13} + B_{34} \, , \qquad
    B_{12} \, B_{24} = B_{13} \, B_{34} 
\ee
(see \cite{DMH01bt}). 
This determines algebraically a forth solution from a given (first) solution 
and two B\"acklund descendants of it (with different parameters).

\section{Toward exact solutions of the dispersionless ncpKP hierarchy}
\label{section:CHdisplimit}
In this section we start with a result that determines a large class 
of exact solutions of an ncpKP hierarchy and use the scaling limit 
toward the dispersionless hierarchy in order to obtain from it a corresponding 
result that determines exact solutions of the latter, which is a pdCM  
hierarchy. Let us recall theorem~4.1 from \cite{DMH07Burgers}.

\begin{theorem}
\label{theorem:KPCH}
Let $(\mathcal{A},\cdot)$ be the algebra of $M \times N$ matrices of functions 
of $\bt$ with the product 
\be
        A \cdot B = A Q B \, ,  \label{Qproduct}
\ee
where the ordinary matrix product is used on the right hand side, and 
$Q$ is a constant $N \times M$ matrix. 
Let $\tilde{\mathcal{X}}$ be an invertible $N \times N$ matrix and 
$\tilde{\mathcal{Y}} \in \mathcal{A}$, such that 
$\tilde{\mathcal{X}}, \tilde{\mathcal{Y}}$ solve the linear heat 
hierarchy (i.e. $\pa_{t_n}(\tilde{\mathcal{X}}) = \pa_{t_1}^n(\tilde{\mathcal{X}})$, $n=2,3,\ldots$, and correspondingly for $\tilde{\mathcal{Y}}$) and satisfy
\be
    \tilde{\mathcal{X}}_{t_1} 
  = R \, \tilde{\mathcal{X}} + Q \, \tilde{\mathcal{Y}} \, , \label{tX_t1}
\ee
with a constant $N \times N$ matrix $R$. The pKP hierarchy in $(\mathcal{A},\cdot)$ 
is then solved by
\be 
      \phi := \tilde{\mathcal{Y}} \, \tilde{\mathcal{X}}^{-1} \; . 
\ee
\end{theorem}
\hfill $\square$

A functional representation of the heat hierarchy condition is 
\be
    \la^{-1}(\tilde{\mathcal{X}} - \tilde{\mathcal{X}}_{-[\la]}) 
  = \tilde{\mathcal{X}}_{t_1} \, ,  \label{tX-heat-hier}
\ee
and correspondingly for $\tilde{\mathcal{Y}}$ (with an indeterminate 
$\la$). 
The theorem provides us with a method to construct exact solutions 
of the ncpKP hierarchy in $(\mathcal{A},\cdot)$. 
The idea is now to take the dispersionless limit of 
(\ref{tX_t1}) and (\ref{tX-heat-hier}). This should then result in conditions 
that determine exact solutions of the pdCM hierarchy in 
$(\mathcal{A},\cdot)$. However, assuming for 
$\tilde{\mathcal{X}},\tilde{\mathcal{Y}}$ power series expansions 
in $\epsilon$ with nonvanishing terms of zeroth order, this results in 
too restrictive conditions. The way out is to note that a 
``gauge transformation'' 
\be
    \tilde{\mathcal{X}} = \mathcal{X} \, G \, , \qquad
    \tilde{\mathcal{Y}} = \mathcal{Y} \, G \, ,
\ee
with an $N \times N$ matrix $G$, leaves $\phi$ invariant. 
Choosing 
\be
    G = \exp(\xi(\bt,P)) \, , \qquad  
    \xi(\bt,P) := \sum_{n \geq 1} t_n \, P^n  \, ,
\ee
with a constant $N \times N$ matrix $P$, and using 
$\xi(\bt,P)_{-[\la]} = \xi(\bt,P) + \ln(\mathcal{I}_N-\la P)$ 
with the $N \times N$ unit matrix $\mathcal{I}_N$, the heat hierarchy 
equations are mapped to 
\be
    (\mathcal{X} - \mathcal{X}_{-[\la]})(\la^{-1}-P)
  = \mathcal{X}_{t_1} \, , \quad
    (\mathcal{Y} - \mathcal{Y}_{-[\la]})(\la^{-1}-P) 
  = \mathcal{Y}_{t_1} \, ,    \label{XY-heat-hier}
\ee
and (\ref{tX_t1}) is converted into
\be
   \mathcal{X}_{t_1} + \mathcal{X} P 
  = R \mathcal{X} + Q \mathcal{Y} \; .  \label{X_t1}
\ee
Assuming that
\be
   \mathcal{X}(\bt,\epsilon) 
 = \mathcal{X}_{(0)}(\bT) + \mathcal{O}(\epsilon) \, , \quad
   \mathcal{Y}(\bt,\epsilon) 
 = \mathcal{Y}_{(0)}(\bT) + \mathcal{O}(\epsilon) \, ,  
\ee
\be
     R = R_{(0)} + \mathcal{O}(\epsilon) \, , \quad
     Q = Q_{(0)} + \mathcal{O}(\epsilon) \, ,
\ee
and $P$ independent of $\epsilon$, then we obtain from (\ref{X_t1})  
\be
    \epsilon \mathcal{X}_{(0),T_1} + \mathcal{X}_{(0)} P 
  = R_{(0)} \mathcal{X}_{(0)} + Q_{(0)} \mathcal{Y}_{(0)} 
    + \mathcal{O}(\epsilon) \, ,
\ee
and from (\ref{XY-heat-hier}) 
\be
  \epsilon \, D(\la) \mathcal{X}_{(0)} \, (\la^{-1}-P)
  = \epsilon \mathcal{X}_{(0),T_1} + O(\epsilon^2) \, ,
\ee
together with the same equation for $\mathcal{Y}_{(0)}$. After dividing the last 
equation by $\epsilon$, these equations have the dispersionless limits
\be
    R_{(0)} \mathcal{X}_{(0)} + Q_{(0)} \mathcal{Y}_{(0)} = \mathcal{X}_{(0)} P \, ,
\ee
respectively
\be
   D(\la) \mathcal{X}_{(0)} \, (\la^{-1}-P) = \mathcal{X}_{(0),T_1} \, , \quad 
   D(\la) \mathcal{Y}_{(0)} \, (\la^{-1}-P) = \mathcal{Y}_{(0),T_1} \, ,
\ee
which is
\be
  \frac{1}{n+1} \mathcal{X}_{(0),T_{n+1}} = \frac{1}{n} \mathcal{X}_{(0),T_n} P \, ,
  \quad 
  \frac{1}{n+1} \mathcal{Y}_{(0),T_{n+1}} = \frac{1}{n} \mathcal{Y}_{(0),T_n} P \, ,
\ee
($n=1,2,\ldots$), or in terms of the variables (\ref{x_n-T_n}),
\be
   \mathcal{X}_{(0),x_{n+1}} = \mathcal{X}_{(0),x_n} P \, ,
   \quad 
   \mathcal{Y}_{(0),x_{n+1}} = \mathcal{Y}_{(0),x_n} P \, ,
   \quad \quad n=1,2,\ldots \; .
\ee
Under the stated conditions, we have an expansion 
\be
    \phi(\bt,\epsilon) = \Phi(x_1,x_2,\ldots) + \mathcal{O}(\epsilon) \, , 
\ee
which determines an exact solution $\Phi$ of the dispersionless limit 
of the ncpKP hierarchy, i.e. the pdCM hierarchy (\ref{disp-ncpKPhier}). 
Proposition~\ref{prop:pdcm_sols} in the following section confirms this directly, 
i.e. without reference to the scaling limit procedure applied to the ncpKP 
hierarchy and the above theorem.

\section{Exact solutions of the pdCM hierarchy}
\label{section:pdc-sols}
The main result of the preceding section will be formulated in the next 
proposition, and we provide a direct proof. It will then be further 
elaborated and applied in order to construct some classes of exact solutions 
of the ($su(m)$) pdCM hierarchy. In this section, 
symbols like $\mathcal{X}$ and $Q$, for example, correspond to 
$\mathcal{X}_{(0)}$ and $Q_{(0)}$ in the preceding sections. 
Since now we resolve our considerations from the dispersionless 
limit procedure, there is no need to carry these indices with us any more. 
In fact, this section can be accessed almost completely without 
reference to the previous ones.

\begin{proposition}
\label{prop:pdcm_sols}
Let $\mathcal{X}$ be an invertible $N \times N$ and $\mathcal{Y}$ 
an $M \times N$ matrix such that 
\be
     R \mathcal{X} + Q \mathcal{Y} = \mathcal{X} P 
                     \label{RX+QY=XP}
\ee
and
\be
      \mathcal{X}_{x_{n+1}} = \mathcal{X}_{x_1} P^n \, , \quad
      \mathcal{Y}_{x_{n+1}} = \mathcal{Y}_{x_1} P^n \, , \qquad
       n=1,2,\ldots \, ,                 \label{XY-lin-hier}
\ee
with constant matrices $P,R$ of size $N \times N$, 
and $Q$ of size $N \times M$. Then 
\be
      \Phi = \mathcal{Y} \mathcal{X}^{-1} 
\ee
solves the pdCM hierarchy 
\be
    \Phi_{x_{n+1},x_m} - \Phi_{x_{m+1},x_n}
  =  [ \Phi_{x_n} , \Phi_{x_m} ]_Q  \qquad\quad m,n=1,2,\ldots 
      \label{Q-pdc-hier}
\ee
(which is (\ref{disp-ncpKPhier}) with $Q_{(0)}$ replaced by $Q$). 
\end{proposition}
{\em Proof:} (\ref{XY-lin-hier}) is equivalent to
\bez
   \mathcal{X}_{x_{n+1}} = \mathcal{X}_{x_n} P \, , \quad
   \mathcal{Y}_{x_{n+1}} = \mathcal{Y}_{x_n} P \, ,
   \qquad n=1,2,\ldots \; .
\eez
In terms of the maps $\rmd , \bd$, defined in section~\ref{subsec:bidiff}, 
this can be expressed as
\bez
    \bd \mathcal{X} = \rmd \mathcal{X} \, P \, , \qquad  
    \bd \mathcal{Y} = \rmd \mathcal{Y} \, P \; .
\eez
Hence
\bez
      (\rmd \Phi) \, \mathcal{X} \, P + \Phi \, \rmd \mathcal{X} \, P
  =  \rmd \mathcal{Y} \, P 
  = \bd \mathcal{Y}
  = (\bd \Phi) \, \mathcal{X} + \Phi \, \bd \mathcal{X} 
  = (\bd \Phi) \, \mathcal{X} + \Phi \, \rmd \mathcal{X} \, P \, , 
\eez
and thus\footnote{We note that this equation can be written as 
$\bd \Phi - (\rmd \Phi) Q \Phi = \rmd \Theta$ with $\Theta := \Phi R$, 
which is (\ref{Phi-Theta-hier}). }
\bez
     \bd \Phi = (\rmd \Phi) \, W \, , 
\eez
where
\bez
    W := \mathcal{X} P \mathcal{X}^{-1} = Q \, \Phi + R \, , 
\eez 
using (\ref{RX+QY=XP}). Since $\rmd$ and $\bd$ satisfy (\ref{bicomplex}), 
and since $Q$ and $R$ are constant, we obtain
\bez
   \bd \rmd \Phi = - \rmd \bd \Phi = (\rmd \Phi) \wedge \rmd W 
               = (\rmd \Phi) \wedge Q \, \rmd \Phi \, , 
\eez
which is the hierarchy (\ref{Q-pdc-hier}). 
\hfill $\square$
\vspace{.1cm}

The next result shows how to obtain via proposition~\ref{prop:pdcm_sols} 
solutions of the pdCM hierarchy in the algebra of $m \times m$ matrices 
with the \emph{usual} matrix product (i.e. without the modification by a 
matrix $Q$ different from the unit matrix). If $\tr(R)=\tr(P)$, 
these solutions have values in $sl(m,\mathbb{C})$.

\begin{proposition} 
\label{prop:trace}
Let $U,V$ be $N \times m$ matrices and 
\be
    \varphi = U^\dagger \Phi V \, , \qquad
     Q = V U^\dagger \, ,  \label{varphi=UPhiV,Q=VU}
\ee
where $\Phi = \mathcal{Y} \mathcal{X}^{-1}$ with 
$\mathcal{X}, \mathcal{Y}$ solving (\ref{RX+QY=XP}). Then 
\be
     \tr(\varphi)  = \tr(P) - \tr(R) \; .
              \label{trvarphi}
\ee 
Under the conditions of proposition~\ref{prop:pdcm_sols}, and if 
$U$ and $V$ are constant, $\varphi$ solves the pdCM hierarchy 
(\ref{varphi_hier}). 
\end{proposition}
{\em Proof:} We have
\bez
    \tr(\varphi) = \tr(U^\dagger \mathcal{Y} \mathcal{X}^{-1} V) 
                 = \tr(V U^\dagger \mathcal{Y} \mathcal{X}^{-1}) 
                 =  \tr( Q \mathcal{Y} \mathcal{X}^{-1}) \; .
\eez
Using (\ref{RX+QY=XP}), this can be rewritten as 
\bez
    \tr(\varphi)  = \tr( \mathcal{X} P \mathcal{X}^{-1} - R ) \, , 
\eez
which is (\ref{trvarphi}). The last statement of the proposition is easily 
verified (see also section~\ref{section:disp-ncpKPhier}).
\hfill $\square$
\vspace{.1cm}

It is helpful to extend (\ref{RX+QY=XP}) to  
\be
    H \mathcal{Z} = \mathcal{Z} P  \, ,  \label{HZ=ZP}
\ee
where
\be
    \mathcal{Z} = \left( \begin{array}{c} \mathcal{X} \\
                  \mathcal{Y} \end{array}\right) \, , \qquad
    H = \left(\begin{array}{cc} R & Q \\ S & L
                \end{array}\right) \, ,     \label{Z,H}
\ee
with constant matrices $L,S$. Indeed, the upper component of (\ref{HZ=ZP}) 
reproduces (\ref{RX+QY=XP}). But now we have an additional equation, 
namely $S \mathcal{X} + L \mathcal{Y} = \mathcal{Y} P$ (which together 
with (\ref{RX+QY=XP}) implies the algebraic Riccati equation 
$S + L \Phi - \Phi R - \Phi Q \Phi =0$ for $\Phi$).  
Although the latter appears to impose an unnecessary restriction, it will 
be helpful in order to determine interesting classes of exact solutions. 
The two equations (\ref{XY-lin-hier}) can be combined into
\be
    \mathcal{Z}_{x_{n+1}} = \mathcal{Z}_{x_1} \, P^n  \qquad \quad
    n=1,2, \ldots \; .    \label{Z-lin-hier}
\ee

Obviously, a transformation 
\be
    \mathcal{Z} = \Gamma \mathcal{Z}' \, , \qquad  
    H = \Gamma H' \Gamma^{-1} \, ,  \label{Gamma-transf}
\ee
with a constant matrix 
\be
    \Gamma = \left(\begin{array}{cc} \Gamma_{11} & \Gamma_{12} \\ 
                \Gamma_{21} & \Gamma_{22} 
                \end{array}\right) \, ,
\ee
preserves the form of the equations (\ref{HZ=ZP}) and (\ref{Z-lin-hier}) 
with the same $P$. 
Consequently, if $\mathcal{Z}'$ solves (\ref{HZ=ZP}) and (\ref{Z-lin-hier}) 
with $H'$, and hence $\Phi' = \mathcal{Y}' \mathcal{X}'^{-1}$ solves 
the pdCM hierarchy with $Q'$, then $\mathcal{Z}$ solves the corresponding 
equations with $H$, and according to proposition~\ref{prop:pdcm_sols} 
\be
  \Phi = \mathcal{Y} \mathcal{X}^{-1} 
       = (\Gamma_{21} + \Gamma_{22} \, \Phi')
          (\Gamma_{11} + \Gamma_{12} \, \Phi')^{-1}  \label{Phi_transf}
\ee
solves the pdCM hierarchy with $Q$.\footnote{We note that the transformation 
(\ref{Phi-gaugetransf}) corresponds to the block-diagonal choice 
$\Gamma = \mathrm{diag}(g_2,g_1^{-1})$. Such a transformation does not 
change a solution $\Phi$ in an essential way. } 
Such a transformation thus relates solutions of different versions of 
the pdCM hierarchy, i.e. with different $Q$ (which means 
different products). Since $Q$ and $Q'$ may have different rank, 
via (\ref{varphi=UPhiV,Q=VU}) one obtains corresponding 
solutions of a pdCM hierarchy in a different matrix algebra. 
An extreme case is $Q'=0$. Then the hierarchy (\ref{Q-pdc-hier}) 
reduces to the system of \emph{linear} equations
\be
    \Phi'_{x_m x_{n+1}} - \Phi'_{x_n x_{m+1}} = 0 
    \qquad m,n=1,2,\ldots \; .    \label{pdcm_linhier}
\ee
The above observation now suggests to first construct a solution $\Phi'$ 
of these linear equations, and then use such a transformation 
(as a ``dressing transformation'') to generate a solution of a 
nonlinear hierarchy.

\begin{proposition}
\label{prop:Phi_transf}
Let $P,L,R$ be constant $N \times N$ matrices. 
Let $\mathcal{X}', \mathcal{Y}'$ solve (\ref{XY-lin-hier}) (which is 
(\ref{Z-lin-hier})) and (\ref{HZ=ZP}) with\footnote{If $\mathcal{X}'$ 
is invertible, then $\Phi' = \mathcal{Y}' \mathcal{X}'^{-1}$ solves 
the linear hierarchy (\ref{pdcm_linhier}). This follows from 
proposition~\ref{prop:pdcm_sols}, since $Q'=0$.  }
\be
       H' = \left(\begin{array}{cc} R & 0 \\ 0 & L
                  \end{array}\right) \; .   \label{H'}
\ee 
Then 
\be
    \Phi = \mathcal{Y}' \, (\mathcal{X}' - K \, \mathcal{Y}')^{-1} 
              \label{Phi-K-formula}
\ee
with any constant $N \times M$ matrix $K$, provided that the inverse 
in (\ref{Phi-K-formula}) exists, solves the pdCM hierarchy (\ref{Q-pdc-hier}) 
with 
\be
       Q = R K - K L \; .  \label{Q=RK-KL}
\ee
\end{proposition}
{\em Proof:} 
Choosing in (\ref{Gamma-transf}) the transformation matrix 
\bez
  \Gamma = \left(\begin{array}{cc} \mathcal{I}_N & -K \\ 0 & \mathcal{I}_M 
                 \end{array}\right) \, , 
\eez
where $\mathcal{I}_N$ is the $N \times N$ unit matrix, we have 
\bez
     H = \Gamma H' \Gamma^{-1} 
       = \left(\begin{array}{cc} R & R K - K L \\ 
                                 0 & L 
               \end{array}\right)  \, , 
\eez
and hence $Q = R K - K L$. Since $\mathcal{Z} = \Gamma \mathcal{Z}'$ again 
satisfies (\ref{HZ=ZP}) and (\ref{Z-lin-hier}), proposition~\ref{prop:pdcm_sols} 
tells us that $\Phi$ given by (\ref{Phi_transf}), which is (\ref{Phi-K-formula}), 
solves (\ref{Q-pdc-hier}) with $Q$ given by (\ref{Q=RK-KL}). 
\hfill $\square$
\vspace{.1cm}

A special case of proposition~\ref{prop:Phi_transf} is formulated next. 
This will turn out to be particularly useful in the following. 

\begin{corollary}
\label{cor:Phi-K-formula}
Let $P,K$ be constant $N \times N$ matrices, and $\mathcal{X}'$ an 
$N \times N$ matrix solution of 
\be
     \mathcal{X}'_{x_{n+1}} = \mathcal{X}'_{x_1} \, P^n 
     \qquad \quad n=1,2, \ldots \, ,  \label{X'-lin-hier}
\ee
such that 
\be
     [ P , \mathcal{X}' ] = 0 \; .  \label{commPX'}
\ee
Then 
\be
    \Phi = (\mathcal{X}' - K)^{-1} \, ,   \label{Phi-K-formula2}
\ee
provided that the inverse exists, solves the pdCM hierarchy with 
$Q$ given by 
\be
     Q = [P , K] \; .     \label{Q=[P,K]}
\ee
If moreover (\ref{varphi=UPhiV,Q=VU}) holds, then 
$\varphi$ solves the pdCM hierarchy (\ref{varphi_hier}) in $sl(m,\mathbb{C})$. 
\end{corollary}
{\em Proof:} We check that the assumptions of this corollary 
constitute a special case of those of proposition~\ref{prop:Phi_transf}. 
(\ref{HZ=ZP}) decomposes into 
\bez
   R \mathcal{X}' = \mathcal{X}' P \, , \qquad
   L \mathcal{Y}' = \mathcal{Y}' P \; .   \label{RX'=X'P}
\eez
Choosing 
\bez
     R = L = P  \, ,  \qquad  \mathcal{Y}' = \mathcal{I}_N \, , 
\eez
this reduces to (\ref{commPX'}), and (\ref{XY-lin-hier}) reduces 
to (\ref{X'-lin-hier}). Since $\Phi'^{-1} = \mathcal{X}'$, 
(\ref{Phi-K-formula}) becomes (\ref{Phi-K-formula2}), and 
(\ref{Q=RK-KL}) becomes (\ref{Q=[P,K]}). 
As a consequence of $R=P$ and proposition~\ref{prop:trace}, 
$\varphi$ has vanishing trace, hence takes values in $sl(m,\mathbb{C})$. 
\hfill $\square$
\vspace{.1cm}

\noindent
\textbf{Example.} Let us choose
\be
  P = \left(\begin{array}{cc} 
          p_1 & 0  \\ 
           0  & p_2 
        \end{array}\right) \, , \quad
  \mathcal{X}' = \left(\begin{array}{cc} 
          f_1 & 0  \\ 
           0  & f_2 
        \end{array}\right) \, , \quad
  Q = \left(\begin{array}{cc} 
           0 & 1 \\ 
           1 & 0 
        \end{array}\right) \, ,
\ee
with real constants $p_1 \neq p_2$ and real functions $f_i$. Then (\ref{commPX'}) 
holds and (\ref{X'-lin-hier}) requires that the function $f_i$ depends 
on the variables $x_1,x_2,\ldots$ only through the combination 
$\omega_i = \sum_{n \geq 1} p_i^{n-1} x_n$. (\ref{Q=[P,K]}) is solved by 
\be
  K = \left(\begin{array}{cc} 
           0 & (p_1-p_2)^{-1} \\ 
           (p_2-p_1)^{-1} & 0 
        \end{array}\right) \; .
\ee
A diagonal part of $K$ can be absorbed in (\ref{Phi-K-formula2}) 
by redefinition of $f_1,f_2$. We obtain
\be
    \Phi = \frac{1}{\mathcal{D}} \left(\begin{array}{cc} 
             f_2 & (p_2-p_1)^{-1} \\ 
             (p_1-p_2)^{-1} & f_1
        \end{array}\right) \, , 
\ee
where
\be
     \mathcal{D} = f_1 f_2 + (p_1-p_2)^{-2} \, , 
\ee
and then the following solution of the $sl(2,\mathbb{R})$ pdCM hierarchy:
\be
   \varphi = \Phi Q 
           = \frac{1}{\mathcal{D}} \left(\begin{array}{cc} 
             (p_2-p_1)^{-1} & f_2  \\ 
              f_1 & (p_1-p_2)^{-1} 
        \end{array}\right)  \; .
\ee 
The corresponding conserved density $\mathcal{E}$ is given by
\be
    \mathcal{E} = - \frac{1+p_1 p_2}{(f_1 f_2 + (p_1-p_2)^{-2})^2} \, 
       \frac{d f_1}{d \omega_1} \, \frac{d f_2}{d \omega_2} \, , 
\ee 
which can take both signs, depending on the values of the parameters. 
Choosing
\be
    f_i = \exp(q_i \, \omega_i) + c_i  \qquad \quad i=1,2 \, ,
\ee
with non-negative constants $c_i$ and real constants $q_i \neq 0$, 
the solution is regular (for all $x_1,x_2,\ldots$). 
For positive $c_i$, $\mathcal{E}$ is exponentially localized, a sort of 
soliton. The first derivatives of the components of $\varphi$ 
are \emph{not} localized, however. 
If $c_1$ or $c_2$ tends to zero, it stretches into a half-infinitely extended 
``line soliton'', the location of which is determined by 
$q_1 \, \omega_1 + q_2 \, \omega_2 =0$. 
\hfill $\square$
\vspace{.1cm}

As pointed out in section~\ref{subsec:dncpKP_prop}, the case where 
$\varphi$ given by (\ref{varphi=UPhiV,Q=VU}) has values in the Lie 
algebra of a unitary group is distinguished by the fact that there is 
a non-negative ``energy'' functional, with density given by $\mathcal{E}$ 
defined in (\ref{energy}). We will therefore concentrate on this case 
in the following. 
We further restrict our considerations to the case $M=N$, hence  
$\mathcal{X}, \mathcal{Y}, \Phi$ are all $N \times N$ matrices. 
Let $U$ and $V$ be constant $N \times m$ matrices. 
If $\Phi$ has the property
\be
    \Phi^\dagger = T \Phi T^{-1}     \label{Phidagg}
\ee
with a constant invertible $N \times N$ matrix $T$ which is anti-Hermitian, 
i.e. $T^\dagger = - T$, then by setting 
\be
     U = T \, V
\ee
we achieve that $\varphi = U^\dagger \Phi V$ is \emph{anti-Hermitian}, 
i.e. 
\be
    \varphi^\dagger = - \varphi \; .
\ee
As a consequence of these conditions, we have 
\be
     \varphi = - V^\dagger T \Phi V \, ,       \label{varphi-VTPhiV}
\ee
and 
\be
    Q = V U^\dagger = - V V^\dagger T \, ,  \label{Q_UV_su} 
\ee
which has the property 
\be
         Q^\dagger = - T Q T^{-1} \; .   \label{Q-su-cond}
\ee 
We note that $V \mapsto V \, \sigma$, with a constant unitary 
$m \times m$ matrix $\sigma$, leaves $Q$ invariant and induces a gauge 
transformation $\varphi \mapsto \sigma^\dagger \, \varphi \, \sigma$. 
This can be used to reduce the freedom in the choice of $V$.

In the following we address exact solutions of the $su(m)$ pdCM 
hierarchy by using the recipe of corollary~\ref{cor:Phi-K-formula}. 
Accordingly we should arrange that the solution $\mathcal{X}'$ 
of the linear hierarchy (\ref{X'-lin-hier}) satisfies
\be
    \mathcal{X}'^\dagger = T \mathcal{X}' T^{-1} \; .  \label{X'dagg}
\ee
If also 
\be
     K^\dagger = T K T^{-1} \, ,   \label{K-su-cond}
\ee
then $\Phi$ given by (\ref{Phi-K-formula2}) satisfies the same relation, 
i.e. (\ref{Phidagg}). 
As a further consequence, (\ref{varphi-VTPhiV}) is then 
anti-Hermitian.

Together with (\ref{X'dagg}), (\ref{commPX'}) implies 
$[T^{-1} P^\dagger T, \mathcal{X}']=0$, 
which is identically satisfied as a consequence of (\ref{commPX'}) if 
$P$ has the property 
\be
    P^\dagger = T P T^{-1} \; .  \label{P-su-cond}
\ee
We note that (\ref{Q=[P,K]}) is consistent with (\ref{Q-su-cond}), 
(\ref{K-su-cond}) and (\ref{P-su-cond}). 
Basically the problem of constructing solutions of (\ref{varphi_hier}) 
in $su(m)$ (on the basis of corollary~\ref{cor:Phi-K-formula}) is 
reduced to the problem of satisfying the algebraic equation 
(\ref{Q=[P,K]}) with $Q$ given by (\ref{Q_UV_su}). 
We summarize our results.

\begin{proposition}
\label{prop:su(m)-sols}
Let $(P,\mathcal{X}',T,V)$ be data consisting of a constant 
$N \times N$ matrix $P$, an $N \times N$ matrix $\mathcal{X}'$, which 
solves (\ref{X'-lin-hier}) and (\ref{commPX'}), a constant anti-Hermitian 
$N \times N$ matrix $T$, and a constant $N \times m$ matrix $V$. 
Furthermore, let (\ref{X'dagg}) and (\ref{P-su-cond}) be satisfied, $Q$ be 
defined by (\ref{Q_UV_su}), and suppose that a solution $K$ of (\ref{Q=[P,K]}) 
and (\ref{K-su-cond}) exists. Then $\varphi = - V^\dagger T \Phi V$, 
with $\Phi$ given by (\ref{Phi-K-formula2}), is a solution of the pdCM 
hierarchy (\ref{varphi_hier}) in the Lie algebra $su(m)$. 
\hfill $\square$
\end{proposition}
\vspace{.1cm}

By application of proposition~\ref{prop:su(m)-sols}, some classes of 
exact solutions of the $su(m)$ pdCM hierarchy will 
be derived in the following subsections. Examples are worked out for the 
$su(2)$ case. Corresponding plots are restricted to the three variables entering 
the first hierarchy equation, and we will always use the coordinates 
$t,x,y$ related to the variables $x_1,x_2,x_3$ by the transformation 
(\ref{cord_xyt}). This is mainly done in order to ease a comparison 
with solutions of Ward's modified chiral model 
(cf. section~\ref{subsec:dncpKP_Ward}).

\subsection{A class of solutions of the $su(m)$ pdCM hierarchy}
\label{subsec:class1}
Assuming that $P$ is diagonal, i.e.
\be
    P = \mathrm{diag}(p_1, \ldots, p_N) \, , 
\ee
with complex constants $p_i \neq p_j$ for $i \neq j$, (\ref{commPX'}) 
requires $\mathcal{X}'$ to be diagonal. 
Writing
\be
    \mathcal{X}' = \mathrm{diag}(f_1,\ldots,f_N) \, , 
\ee
where the entries are functions of $x_1,x_2,\ldots$, (\ref{X'-lin-hier}) 
becomes 
\be
     f_{j,x_n} = p_j^{n-1} f_{j,x_1} 
     \qquad \quad j=1,\ldots,N, \quad n=1,2,\ldots \; . 
\ee
This is solved if $f_j$ is a \emph{holomorphic}\footnote{More generally, 
the function $f_j$ is allowed to have singularities in the complex 
$\omega_j$-plane, but we will not consider such solutions in this work. 
See also e.g. \cite{Ioan+Mant05} in the case of Ward's chiral model. }
function of 
\be
    \omega_j := \sum_{n \geq 1} x_n \, p_j^{n-1}   \label{omega_j}
\ee
(with the same $j$ both for the function and its argument).
In particular, $f_j$ depends on the variables $x_1,x_2,\ldots$ only 
through the combination (\ref{omega_j}). 
The condition (\ref{X'dagg}) with an invertible anti-Hermitian matrix 
$T$ imposes restrictions on the set of functions $\{ f_j \}_{j=1,\ldots,N}$, 
see section~\ref{subsec:regsols}.

According to corollary~\ref{cor:Phi-K-formula}, $\Phi$ given by 
(\ref{Phi-K-formula2}) solves the pdCM hierarchy with $Q$ given by 
(\ref{Q=[P,K]}), which implies $Q_{ii}=0$, $i=1,\ldots,N$, and
\be
    K_{ij} = \frac{Q_{ij}}{p_i-p_j} 
           = - \sum_{a=1}^m \sum_{k=1}^N \frac{V_{ia} V^\ast_{ka} T_{kj}}{p_i-p_j} 
               \qquad\quad   i \neq j \, , 
\ee
where we took (\ref{Q_UV_su}) into account. (\ref{Phi-K-formula2}) shows that 
a diagonal part of $K$ can be absorbed by redefinition of the functions $f_j$.  
Hence it is no restriction to assume that $K_{ii}=0$ for $i=1,\ldots, N$. 
(\ref{K-su-cond}) is then satisfied as a consequence of (\ref{P-su-cond}). 
Now (\ref{Phi-K-formula2}) can be expressed as
\be
    (\Phi^{-1} )_{ij} = f_i \, \delta_{ij} 
  + \sum_{k=1}^N \sum_{a=1}^m \frac{V_{ia} V^\ast_{ka} T_{kj}}{p_i-p_j} \, , 
\ee
and $\varphi = - V^\dagger T \Phi V$ solves the $su(m)$ pdCM hierarchy 
(\ref{varphi_hier}).

\subsubsection{Some regular and localized solutions of the $su(2)$ pdCM hierarchy}
\label{subsec:regsols}
Choosing $N$ even and for $T$ the following block-diagonal form, 
\be
    T = \left(\begin{array}{ccccc} 
          0 & -1 &        &   &     \\ 
          1 & 0  &        &   &     \\
            &    & \ddots &   &     \\
            &    &        & 0 & -1  \\
            &    &        & 1 & 0
        \end{array}\right) \, ,    \label{class1_T}
\ee
the conditions (\ref{X'dagg}) and (\ref{P-su-cond}) read
\be
  && f_2(\omega_2) = f_1(\omega_1)^\ast, \ldots, 
     f_N(\omega_N) = f_{N-1}(\omega_{N-1})^\ast \, , \\
  && p_2 = p_1^\ast, \ldots, p_N=p_{N-1}^\ast \; .
\ee
We also have $\omega_2 = \omega_1^\ast, \ldots, \omega_N = \omega_{N-1}^\ast$. 
\vspace{.1cm}

\noindent
\textbf{Example~1.} 
Let $N=2$ and $V = \mathcal{I}_2$, which leads to $Q=-T$. Then we obtain
\be
    \varphi = \frac{\beta}{1 + \beta^2 |f(\omega)|^2}
     \left(\begin{array}{cc} 
          -\rmi & \beta \, f(\omega)   \\ 
          - \beta \, f(\omega)^\ast & \rmi  
        \end{array}\right) 
\ee
where $\rmi=\sqrt{-1}$, $\beta = 2 \, \Im(p)$, $p = p_1$, and $f = f_1$ 
is an arbitrary holomorphic function of $\omega = \sum_{n \geq 1} p^{n-1} x_n$. 
This solution is regular for all $x_1,x_2,\ldots$. 
The corresponding ``energy density'' is given by
\be
   \mathcal{E} = \beta^4 \, (1+ |p|^2) \, 
   \left( 1 + \beta^2 |f(\omega)|^2 \right)^{-2} \, 
    \left| \frac{df}{d \omega} \right|^2  \; . 
\ee
Choosing for $f$ a non-constant polynomial in $\omega$, the solution 
is rational and localized, thus a field configuration that is often 
called a ``lump''. The shape of $\mathcal{E}$ 
depends on the degree of the polynomial and in particular on its zeros. 

\begin{figure}[t] 
\begin{center} 
\resizebox{16cm}{!}{
\includegraphics{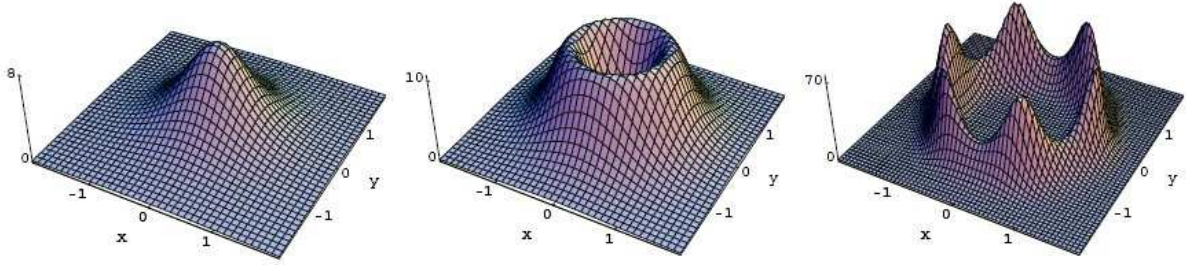}
}
\end{center} 
\caption{Plots of $\mathcal{E}$ for stationary lump solutions of 
the $su(2)$ pdCM equation, according to example~1 of 
section~\ref{subsec:regsols}. Here we chose $p=\rmi$ and 
$f(\omega) = \omega/2$ (left), $f(\omega) = \omega^2/2$ 
(middle) and $f(\omega) = (\omega^6-1)/8$ (right). 
\label{fig:su(2)_1lump}}
\end{figure}

Let $x_4,x_5,\ldots =0$ for the moment, so we concentrate on the first hierarchy 
equation. Applying the coordinate transformation (\ref{cord_xyt}), we have
\be  
     \omega = \frac{1}{2}(t-x+2py+p^2(t+x)) \; . \label{omega}
\ee
We note that this becomes $t$-independent if $p=\pm \rmi$ (i.e. $\beta= \pm 2$), 
in which case $\omega= -x \pm \rmi \, y$, and the solution $\varphi$ 
is stationary. For $f(\omega) = q \, \omega + c$, we obtain a simple lump. 
For example, choosing $p=\rmi$ and $f(\omega) = \omega/2$, we have 
\be
    \mathcal{E} = \frac{8}{(1+x^2+y^2)^2} \, , 
\ee  
see figure~\ref{fig:su(2)_1lump}. $c \neq 0$ causes a displacement of the 
lump in the $xy$-plane. 

For $f(\omega) = q \, (\omega - c)^n$, $n>1$, with a zero of $n$th order, 
$\mathcal{E}$ is bowl-shaped. In particular, if 
$p=\rmi$ and $f(\omega) = \omega^2/2$, we have 
\be
     \mathcal{E} = \frac{32 \, (x^2+y^2)}{(1+(x^2+y^2)^2)^2} \, , 
\ee
which is shown in figure~\ref{fig:su(2)_1lump} (second plot). 
The third plot in figure~\ref{fig:su(2)_1lump} displays another example. 

Configurations with $M$ lumps are obtained by choosing 
$f$ as a product of (powers of) factors $\omega - c_i$, $i=1,\ldots,M$, 
with pairwise different complex constants $c_i$.  
\hfill $\square$

\begin{figure}[t] 
\begin{center} 
\resizebox{16cm}{!}{
\includegraphics{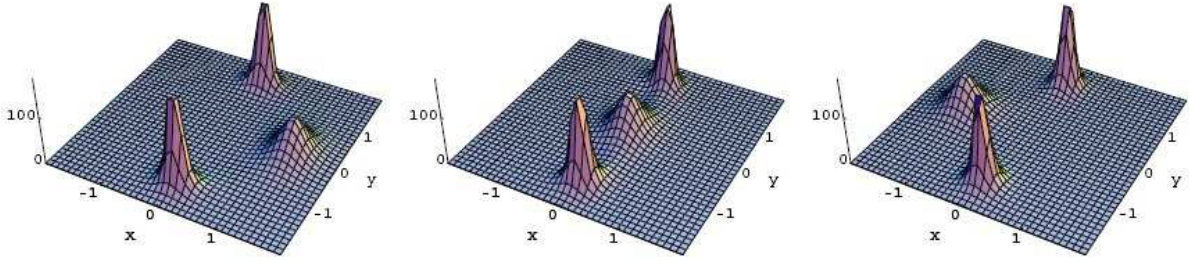}
}
\end{center} 
\caption{Plots of $\mathcal{E}$ at $t=-2,0,2$ for a three lump 
solution of the $su(2)$ pdCM equation, according to example~2 of 
section~\ref{subsec:regsols}. Here we chose $p_1 = \rmi, p_2 = 2 \rmi$, 
and $f_1(\omega_1) = \omega_1^2 +2$, $f_2(\omega_2) = \omega_2/4$. 
Due to the special choice of $p_1$, a pair of lumps is stationary. 
The positions of the latter are given by the zeros of 
$f_1(\omega_1) = x^2 - y^2 +2 - 2 \rmi x y$, which are located at 
$(x,y)=(0,\pm \sqrt{2})$. The position of the third lump corresponds to 
the zero of $f_2(\omega_2)$, which is given by $(x,y)=(-3t/5,0)$. 
Choosing $-2$ instead of $+2$ in $f_1(\omega_1)$, all three lumps are 
located on the line $y=0$, and the third lump moves through both 
members of the pair (which then reside at $x=\pm \sqrt{2}$). 
\label{fig:3nonintlumps} }
\end{figure}

\begin{figure}[t] 
\begin{center} 
\resizebox{12cm}{!}{
\includegraphics{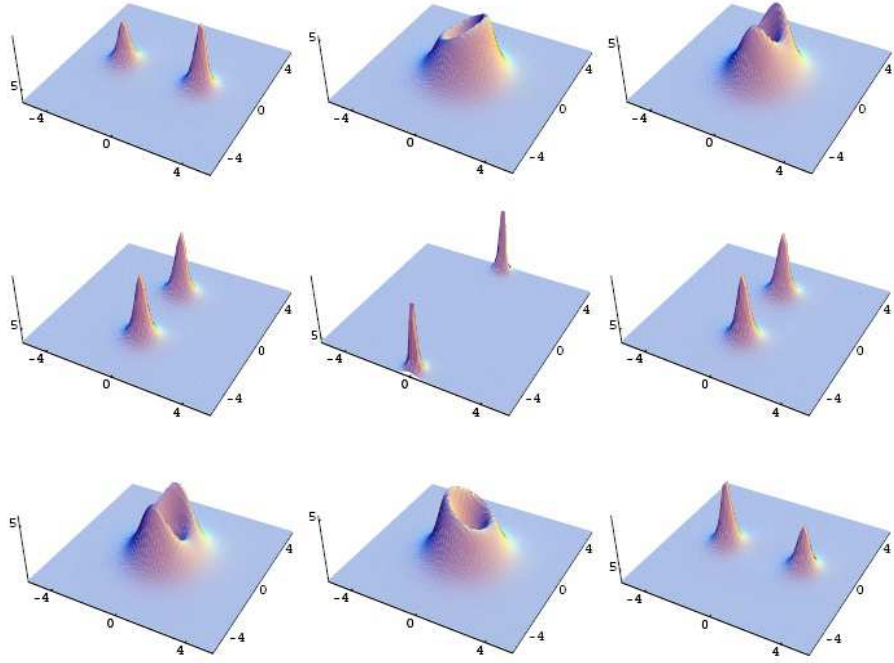}
}
\end{center} 
\caption{Plots of $\mathcal{E}$ at $t=-110,-100,-99,-90,0,90,99,100,110$  
for the solution in example~2 of section~\ref{subsec:regsols} 
with the data $p_1 = -19 \, \rmi/20$, $p_2=21 \, \rmi/20$, 
$f_1 = 2 \, \rmi \, \omega_1$, $f_2 = 2 \, \rmi \, \omega_2$. Two lumps 
approach each other in $x$-direction, merge, move away from one another 
in $y$-direction up to some maximal distance, return to each other and merge 
again, and then separate in $x$-direction. 
\label{fig:2lumps_int} }
\end{figure}

\vspace{.1cm}
\noindent
\textbf{Example~2.} Let $N=4$ and 
\be
   V = \left( \begin{array}{c} \mathcal{I}_2 \\ \mathcal{I}_2
               \end{array} \right) \; .
\ee
Then $\varphi$ has the following components,
\be
 &&  \varphi_{11} = -\varphi_{22} = \frac{1}{\mathcal{D}} \Big( 
     \beta_1 \beta_2 ( a h_1^\ast h_2 - a^\ast h_1 h_2^\ast)
    -\rmi \, [ (\beta_1+\beta_2) |b|^4 + \beta_1 |a h_2|^2 + \beta_2 |a h_1|^2]
    \Big) \, ,  \nonumber \\
 &&  \varphi_{12} = -\varphi_{21}^\ast
    =  \frac{1}{\mathcal{D}} \Big( 
      (b^\ast)^2 (a \beta_1 h_1 + a^\ast \beta_2 h_2) 
      + a |h_1|^2 \beta_2 h_2 + a^\ast \beta_1 h_1 |h_2|^2 \Big) \, ,
\ee
where
\be
 &&   \beta_i = 2 \, \Im(p_i) \, , \quad 
      a = p_1 - p_2^\ast \, , \quad
      b = p_1 - p_2 \, ,               \cr
 &&   h_1 = a \beta_1 f_1 \, , \qquad 
      h_2 = a^\ast \beta_2 f_2  \, ,   \cr 
 &&   \mathcal{D} = (|b|^2 + |h_1|^2)(|b|^2 + |h_2|^2) 
        + \beta_1 \beta_2 |h_1 - h_2|^2 \; .
\ee
This solution is regular since $\mathcal{D}$ is positive (note that 
$|b| >0$ since $p_1 \neq p_2$, $|b|^2 \geq -\beta_1 \beta_2$, and 
use $|h_1|^2 + |h_2|^2 \geq |h_1 - h_2|^2$). 
Figure~\ref{fig:3nonintlumps} shows an example. 
For generic parameter values, plots of $\mathcal{E}$ show lumps with 
apparently trivial interaction. 
But if $p_1,p_2$ are close to the values $\pm \rmi$ (that correspond 
to the stationary single lump solutions), a non-trivial interaction 
is observed in a compact space region, see figure~\ref{fig:2lumps_int}. 
The scalar KP-I equation possesses solutions with the same behaviour 
\cite{LTG04}. Moreover, also dipolar vortices (modons) of a 
barotropic equation \cite{McWi+Zabu82} and BPS monopoles
\cite{Gibb+Mant86} show such a behaviour in head-on collisions. 
\hfill $\square$

\subsection{Another class of solutions of the $su(m)$ pdCM hierarchy}
\label{subsec:class2}
Let $N$ be even. We introduce the commuting matrices 
\be  
  \mathcal{X}_I = \left( \begin{array}{cc} f_I & \tilde{h}_I \\ 
                                    0 & f_I  
                \end{array} \right)
  \, , \quad
  P_I = \left( \begin{array}{cc} p_I & 1 \\ 
                                   0 & p_I  
               \end{array} \right) \, ,  \label{class2:X_I,P_I}
\ee 
with pairwise different complex constants $p_I$, and functions 
$f_I,h_I$, $I=1,\ldots, N/2$, 
and construct in terms of them the block-diagonal matrices
\be
    \mathcal{X}' 
= \left( \begin{array}{cccc} \mathcal{X}_1 &  &  &  \\ 
                                     & \mathcal{X}_2 & &  \\
                                     &  & \ddots &  \\
                                     &  &  & \mathcal{X}_{N/2}
                     \end{array} \right) 
    \, , \quad 
    P = \left( \begin{array}{cccc} P_1 &  &  &  \\ 
                                     & P_2 & &  \\
                                     &  & \ddots &  \\
                                     &  &  & P_{N/2}
                     \end{array} \right)  \, , \label{class2:X',P}
\ee
which then obviously also commute. Now (\ref{X'-lin-hier}) becomes
\be
  f_{I,x_n} = p_I^{n-1} f_{I,x_1} \, , \;  
  \tilde{h}_{I,x_n} = p_I^{n-1} \tilde{h}_{I,x_1} + (n-1) p^{n-2} f_{I,x_1} \, , 
\ee
where $I=1,\ldots,N/2$ and $n=1,2,\ldots$.
Writing
\be
    \tilde{h}_I = h_I + \frac{\pa f}{\pa p_I} \, ,  \label{th_I}
\ee
the second equation is turned into $h_{I,x_n} = p_I^{n-1} h_{I,x_1}$, by 
use of the first. Hence (\ref{X'-lin-hier}) is satisfied if, for $I=1,\ldots,N/2$,  
$f_I$ and $h_I$ are holomorphic functions of
\be
    \omega_I = \sum_{n \geq 1} x_n \, p_I^{n-1}  \label{omega_I}
\ee
(which is (\ref{omega_j})), and in particular only depend on the 
variables $x_1,x_2,\ldots$ through this combination. 
In order to explore the consequences of (\ref{Q=[P,K]}), we write 
$K$ and $Q$ as $N/2 \times N/2$ matrices, where the components 
$K_{IJ}$, respectively $Q_{IJ}$, are $2 \times 2$ matrices.

\begin{proposition}
\label{prop:K_IJ}
With the matrix $P$ defined in (\ref{class2:X',P}), and any $Q$, 
the solution of (\ref{Q=[P,K]}) is given by 
\be
    K_{IJ} = \frac{Q_{IJ}}{p_I-p_J} - \frac{[\Pi_2,Q_{IJ}]}{(p_I-p_J)^2} 
    + \frac{[\Pi_2,[\Pi_2,Q_{IJ}]]}{(p_I-p_J)^3}     \label{K_IJ-sol}
\ee
for $I \neq J$, and 
\be
    [ \Pi_2 , K_{JJ} ] = Q_{JJ}  \qquad\quad J=1,\ldots,N/2  \, , 
            \label{class2:QJJ=[Pi,KJJ]}
\ee
where
\be
    \Pi_2 = \left( \begin{array}{cc} 0 & 1 \\ 
                                     0 & 0  
               \end{array} \right) \; . 
\ee 
\end{proposition}
{\em Proof:} We write $P_I = p_I \, \mathcal{I}_2 + \Pi_2$. Then 
(\ref{Q=[P,K]}), restricted to components with $I \neq J$, takes the form
\bez
   \Big( \id + \frac{1}{p_I - p_J} \, \mathrm{ad}_{\Pi_2} \Big) K_{IJ}
 = \frac{Q_{IJ}}{p_I - p_J} \, , 
\eez
where $\mathrm{ad}_{\Pi} K = [\Pi,K]$. Now (\ref{K_IJ-sol}) follows from  
\bez
    \Big( \id + \frac{1}{p_I - p_J} \, \mathrm{ad}_{\Pi_2} \Big)^{-1} 
  = \sum_{k=0}^2 (-1)^k (p_I - p_J)^{-k} \, \mathrm{ad}_{\Pi_2}^k \, , 
\eez
since $\mathrm{ad}_{\Pi_2}^3 =0$. 
The diagonal components of (\ref{Q=[P,K]}) are 
$\mathrm{ad}_{\Pi_2} K_{JJ} = Q_{JJ}$, which is (\ref{class2:QJJ=[Pi,KJJ]}).
\hfill $\square$
\vspace{.1cm}

\noindent
\textbf{Remark.} 
In view of (\ref{Phi-K-formula2}), we may always assume that 
the two upper entries of $K_{JJ}$ vanish (since non-vanishing entries 
can be absorbed into $\mathcal{X}'$). Using the matrix $T$ given below 
in (\ref{T,tau_2}), the condition (\ref{K-su-cond}) then implies 
$K_{2J-1,2J-1}^\dagger = \tau_2 K_{2J,2J} \tau_2$, $J=1, \ldots, N/2$, 
and this requires that $K_{JJ}$ can only have a non-zero entry in 
the lower left corner. 
As a consequence of (\ref{class2:QJJ=[Pi,KJJ]}), $Q_{JJ}$ is then 
diagonal and has vanishing trace. 
\hfill $\square$
\vspace{.1cm}

A simple way of satisfying (\ref{class2:QJJ=[Pi,KJJ]}) is to choose $V$ 
such that the diagonal blocks $Q_{JJ}$ vanish, and then set $K_{JJ} =0$, 
$J=1,\ldots,N/2$. This will be done in section~\ref{subsec:anom_lumps}.  

It remains to satisfy the further anti-Hermiticity conditions.

\subsubsection{$su(2)$ lumps with ``anomalous'' scattering}
\label{subsec:anom_lumps}
Let $N$ now be a multiple of $4$. In analogy with (\ref{class1_T}) we set
\be
    T = \left(\begin{array}{ccccc} 
             0 & -\tau_2 &        &        &         \\
        \tau_2 &  0      &        &        &         \\
               &         & \ddots &        &         \\
               &         &        &   0    & -\tau_2 \\
               &         &        & \tau_2 & 0 
              \end{array}\right) \quad \mbox{where} \quad
    \tau_2 = \left(\begin{array}{cc} 0&1 \\ 1&0 \end{array}\right) \, .
    \label{T,tau_2}
\ee
Then (\ref{P-su-cond}) means $P_1^\dagger = \tau_2 P_2 \tau_2, \ldots, 
P_{N/2-1}^\dagger = \tau_2 P_{N/2} \tau_2$, which by use of 
(\ref{class2:X_I,P_I}) amounts to 
\be
   p_{2J} = p_{2J-1}^\ast \qquad \quad J=1,\ldots,N/2 \; .
\ee

Since we address the case $m=2$, $V$ has to be chosen as an $N \times 2$ matrix, 
which we subdivide into $2 \times 2$ blocks $V_I$, $I=1,\ldots,N/2$. 
It follows that
\be
    Q_{JJ} = \left\{ \begin{array}{r} 
             - V_J V_{J+1}^\dagger \, \tau_2 \\
               V_J V_{J-1}^\dagger \, \tau_2 
                     \end{array} \quad \mbox{if $J$ is} \quad
             \right. 
                     \begin{array}{l} \mbox{odd} \\ \mbox{even} 
                     \end{array}  \; . 
\ee 
Thus, in order to achieve that $Q_{JJ}=0$, we must arrange that
\be
    V_{2J-1} V_{2J}^\dagger = 0 \qquad \quad   J=1,\ldots, N/2 \; .  
          \label{Q_JJ=0-cond}
\ee
Then $Q$ has the following structure 
\be
   Q = \left(\begin{array}{ccccc} 
      0 & V_1 V_1^\dagger & - V_1 V_4^\dagger & V_1 V_3^\dagger & \cdots  \\
     -V_2 V_2^\dagger & 0 & - V_2 V_4^\dagger & V_2 V_3^\dagger & \cdots  \\
     -V_3 V_2^\dagger & V_3 V_1^\dagger & 0 & V_3 V_3^\dagger & \cdots    \\
     -V_4 V_2^\dagger & V_4 V_1^\dagger & -V_4 V_4^\dagger & 0 &           \\
        \vdots        & \vdots          & \vdots          &   & \ddots 
              \end{array}\right) \, \tau_2  \; . 
\ee
\vspace{.1cm}

\begin{figure}[t] 
\begin{center} 
\resizebox{16cm}{!}{
\includegraphics{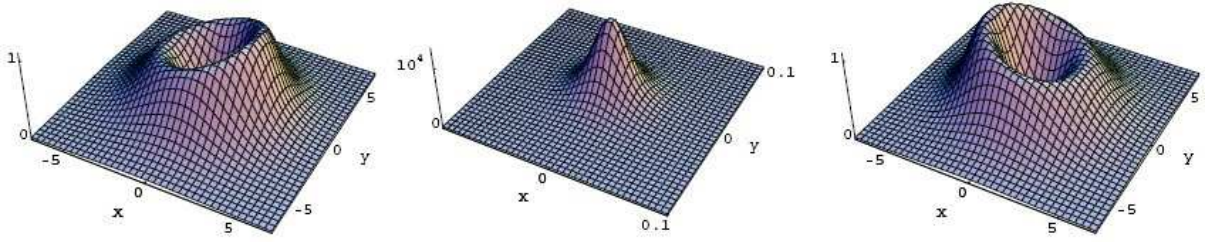}
}
\end{center} 
\caption{Plots of $\mathcal{E}$ at $t=-50,0,50$ for the solution of 
section~\ref{subsec:anom_lumps} with the data 
$p_1 = \rmi$ (i.e. $\beta =2$), $f = 2 \, \omega$ and $h = 0$.  
\label{fig:2Jordan_180degr_rot} }
\end{figure}

\begin{figure}[t] 
\begin{center} 
\resizebox{16cm}{!}{
\includegraphics{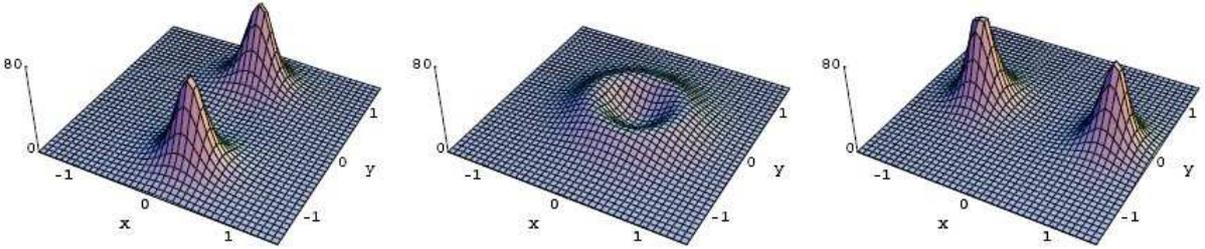}
}
\end{center} 
\caption{Plots of $\mathcal{E}$ at $t=-5,0,5$ for a 2-lump 
solution, exhibiting ``scattering at right angle'', see 
section~\ref{subsec:anom_lumps}. Here we chose 
$p_1 = \rmi$ and $f = -\rmi \, \omega/32$, $h = -\omega^2/4$. 
\label{fig:2Jordan_90degr_scatt} }
\end{figure}

\begin{figure}[t] 
\begin{center} 
\resizebox{16cm}{!}{
\includegraphics{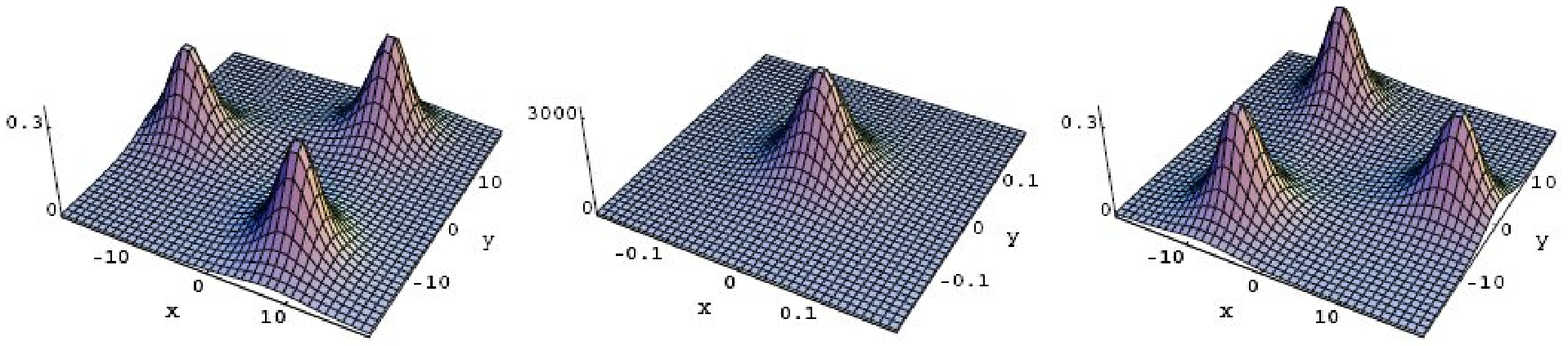}
}
\end{center} 
\caption{Plots of $\mathcal{E}$ at $t=-500,0,500$ for a 3-lump  
configuration exhibiting $60^\circ$ scattering. This is obtained 
from the solution of the example in section~\ref{subsec:anom_lumps}
with the data $p_1 = \rmi$ and $f = -\rmi \, \omega$, $h = \omega^3/8$. 
\label{fig:2Jordan_60degr_scatt} }
\end{figure}

\begin{figure}[t] 
\begin{center} 
\resizebox{16cm}{!}{
\includegraphics{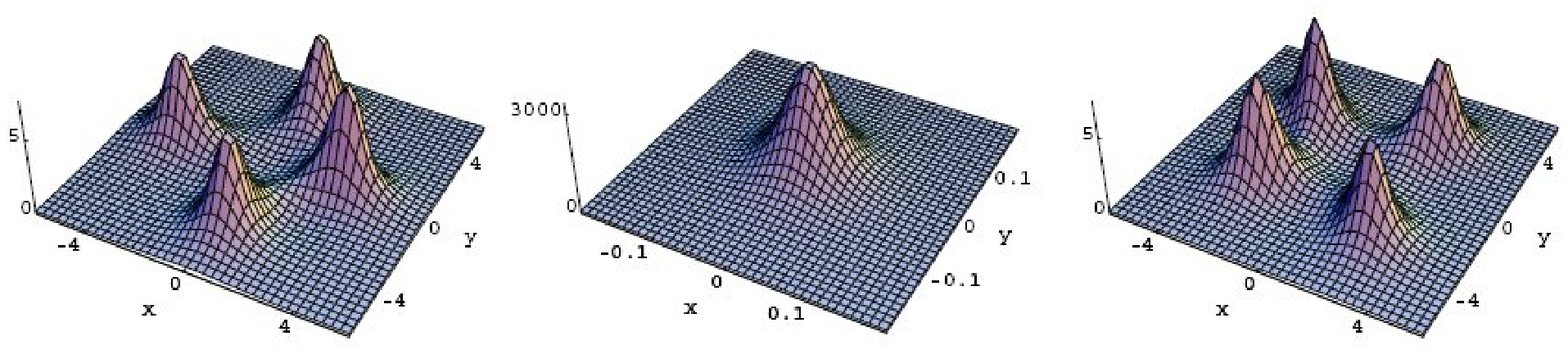}
}
\end{center} 
\caption{Plots of $\mathcal{E}$ at $t=-20,0,20$ for a 4-lump  
configuration exhibiting $45^\circ$ scattering. This is obtained 
from the solution of the example in section~\ref{subsec:anom_lumps}
with the data $p_1 = \rmi$ and $f = -\rmi \, \omega$, $h = \omega^4/8$. 
\label{fig:2Jordan_45degr_scatt} }
\end{figure}

\noindent
\textbf{Example.} The simplest case is $N=4$. Excluding degenerate cases, 
the two blocks $V_1, V_2$ of $V$ should both have rank $1$. Hence
$V_1 = v_1 u_1^\dagger, V_2 = v_2 u_2^\dagger$ with vectors $u_J, v_J$, 
$J=1,2$, satisfying $u_1^\dagger u_2 =0$. With a unitary transformation 
$\sigma$ we can achieve that the lower component of $u_1$ vanishes. 
It follows that the upper component of $u_2$ also vanishes. By a 
redefinition of $v_1,v_2$, we obtain $u_1^\dagger = (1,0)$ and 
$u_2^\dagger = (0,1)$, and thus
\be
 &&   Q_{12} = v_1 (\tau_2 v_1)^\dagger 
           = \left(\begin{array}{cc} 
                v_{11} v_{12}^\ast & |v_{11}|^2   \\
                        |v_{12}|^2 & v_{12} v_{11}^\ast
             \end{array} \right) 
             \, ,  \cr
 &&   Q_{21} = - v_2 (\tau_2 v_2)^\dagger 
           = - \left(\begin{array}{cc} 
                v_{21} v_{22}^\ast & |v_{21}|^2   \\
                        |v_{22}|^2 & v_{22} v_{21}^\ast
             \end{array} \right) \, , 
\ee
with an obvious notation for the components of $v_1$ and $v_2$. 
We should exclude the case when the expression (\ref{K_IJ-sol}) for $K$ 
reduces to the first term on the right hand side, since this leads back 
to the solution of example~1 in section~\ref{subsec:class1}. This case is 
ruled out if $v_{12}$ or $v_{22}$ is different from zero, which suggests 
to choose $v_J =(0,1)$, $J=1,2$, and thus 
\be
   V_1 = \left( \begin{array}{cc} 0 & 0 \\ 
                                  1 & 0 
                \end{array} \right) \, , \qquad
   V_2 = \left( \begin{array}{cc} 0 & 0 \\ 
                                  0 & 1
                  \end{array} \right) \; .   \label{class2:V1,V2}
\ee
Then proposition~\ref{prop:K_IJ} yields 
\be
    K = \left( \begin{array}{cc}  0 & K_{12} \\ -K_{21} & 0  
        \end{array} \right) \, , \label{K_2intlumps}
\ee
where 
\be
    K_{IJ} = \left( \begin{array}{ccc}
             -(p_I-p_J)^{-2} & -2 \, (p_I-p_J)^{-3} \\
             (p_I-p_J)^{-1} & (p_I-p_J)^{-2} 
         \end{array} \right) 
\ee
for $I \neq J$. 
This in turn allows to compute $\Phi$ and then also $\varphi$. 
The anti-Hermiticity conditions are then satisfied by setting 
\be
   p_2 = p_1^\ast \, , \quad f_2(\omega_2) = f_1(\omega_1)^\ast \, , \quad
   h_2(\omega_2) = h_1(\omega_1)^\ast \, ,
\ee
and we have $\omega := \omega_1 = \omega_2^\ast$. The result is
\be
  &&  \varphi_{11} = - \varphi_{22}
  = - \frac{\rmi}{\beta^7 \mathcal{D}} \left( 2 + \beta^4 |f|^2
        + \beta^4 |f + \rmi \, \beta \tilde{h}|^2 \right) \, ,  \cr
  &&  \varphi_{12} = - \varphi_{21}^\ast
  = \frac{1}{\beta^5 \mathcal{D}} \left( 4 \, \rmi \, f - \beta \tilde{h}
     - \beta^5 f^2 \tilde{h}^\ast \right) \, ,
\ee
with $\beta = 2 \Im(p_1)$, $f = f_1(\omega)$, $\tilde{h} = \tilde{h}_1$ given 
by (\ref{th_I}) in terms of $f$ and $h=h_1(\omega)$, and
\be
   \mathcal{D} = \beta^{-8} (1+\beta^4 |f|^2)^2 
                 + \beta^{-4} |2 f + \rmi \, \beta \tilde{h}|^2 \; .
\ee
The solution $\varphi$ is thus regular for any choice of $p_1$, with 
non-vanishing imaginary part, and the holomorphic functions $f,h$. 
An example with $h=0$ is shown in figure~\ref{fig:2Jordan_180degr_rot}. 
More interesting structures appear for non-constant $h$. 
Indeed, figure~\ref{fig:2Jordan_90degr_scatt} shows two lumps that  
scatter at an angle of $90^{\circ}$.
Choosing $f$ linear in $\omega$ and $h$ proportional to $\omega^n$, we 
observe a $\pi/n$ scattering. Figures \ref{fig:2Jordan_60degr_scatt} and 
\ref{fig:2Jordan_45degr_scatt} show examples of $60^\circ$, respectively 
$45^\circ$ scattering. 
\hfill $\square$

\vspace{.1cm} 
Solutions with $\pi/n$ scattering have also been found 
in Ward's chiral model numerically \cite{Sutc92,PSZ92}, 
and analytically as certain limits of families of non-interacting lumps 
\cite{Ward95,Ioan96,Ioan+Zakr98JMP,Dai+Tern07}. 
Moreover, also the scalar KP equation (with positive dispersion, i.e. KP I) 
possesses solutions with this behaviour 
\cite{GPS93JETP,Peli94,ISS95,Ablo+Vill97,Vill+Ablo99,ACTV00}.
In fact, $\pi/n$ scattering in head-on collisions of soliton-like 
objects is a familiar feature of many models (see 
\cite{Rose+Sriv91,KPZ93,MacK95}, in particular). 
It occurs in dipolar vortex collisions \cite{McWi+Zabu82,Nguy+Somm88,vanH+Flor89,Voro+Afan92}, 
in $O(3)$ and $\mathbb{C}\mathbb{P}^1$ models 
\cite{Ward85,Lees90,Lees91,Zakr91,PSZ92,Cova+Zakr97,Spei98,Mant+Sutc04}, 
in Skyrme models \cite{Mant87,LPZ90,Sutc91,KPZ93,Mant+Sutc04}, 
for vortices of the Abelian Higgs (or Ginzburg-Landau) model  
\cite{MMR88,Ruba88,Shel+Ruba88,MRS92,Stra92JMP,Samo92,Mant91,Burz+McCa91,Abde+Burz94,Arth+Burz96,Mant+Sutc04}, 
and BPS monopoles of a $SU(m)$ Yang-Mills-Higgs system 
\cite{Atiy+Hitc85,AHST85,Gibb+Mant86,Danc+Lees93,Mant+Sutc04}. 
Another integrable system that possesses solutions with this behaviour 
is the Davey-Stewartson II equation \cite{Mana+Sant97,Vill+Ablo03} 
(which can actually be obtained by a multiscale expansion from the KP equation 
\cite{Zakh+Kuzn86,AMS90}).

The fact that lumps can interact either trivially or non-trivially 
(in Ward's chiral model) has been attributed to the status of the 
internal degrees of freedom in the solutions \cite{Ward95}. 
But such an explanation appears not to be applicable to the case 
of the scalar KP equation. This requires further clarification.

\subsection{A further generalization}
\label{subsec:general}
In case of the solutions obtained in section~\ref{subsec:class2}, the matrix 
$P$ consists of complex conjugate pairs of $2 \times 2$ blocks of Jordan 
normal form. Of course, this can be generalized to $N_I \times N_I$ 
Jordan blocks
\be
 P_I = \left( \begin{array}{ccccc} 
       p_I & 1      & 0      & \cdots &  0     \\ 
         0 & p_I    & 1      & \ddots & \vdots \\
    \vdots & \ddots & \ddots & \ddots & \vdots \\
    \vdots &        & \ddots &  p_I   &    1   \\
         0 & \cdots & \cdots &  0     & p_I   
       \end{array} \right) \, , 
\ee
and $P$ can be chosen as a block-diagonal matrix with pairs of 
conjugate blocks of this form. For each pair $(P_I,P_I^\ast)$ in $P$, 
the matrix $T$ should then have a corresponding block
\be
 T_I = \left( \begin{array}{cccccc}  &       &   &    &       & -1 \\ 
                                     &       &   &    &\iddots&    \\
                                     &       &   & -1 &       &    \\
                                     &       & 1 &    &       &    \\
                                     &\iddots&   &    &       &    \\
                                   1 &       &   &    &       &
                  \end{array} \right) 
\ee
of size $2N_I \times 2N_I$, in order to achieve that (\ref{P-su-cond}) holds.
\vspace{.1cm}

\begin{figure}[t] 
\begin{center} 
\resizebox{16cm}{!}{
\includegraphics{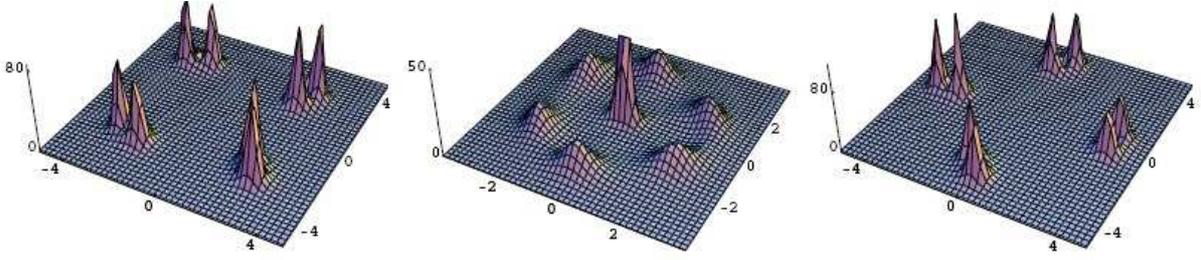}
}
\end{center} 
\caption{Plots of $\mathcal{E}$ at $t=-20,0,20$ for the solution of the 
example in section~\ref{subsec:general} with the data 
$p = \rmi$, $f = -\rmi \, \omega$, $h = \omega^4/8$ and $g = 0$. 
At $t=0$ we have cut off an extremely large lump in the center. 
\label{fig:3Jordan_b} }
\end{figure}

\begin{figure}[t] 
\begin{center} 
\resizebox{16.5cm}{!}{
\includegraphics{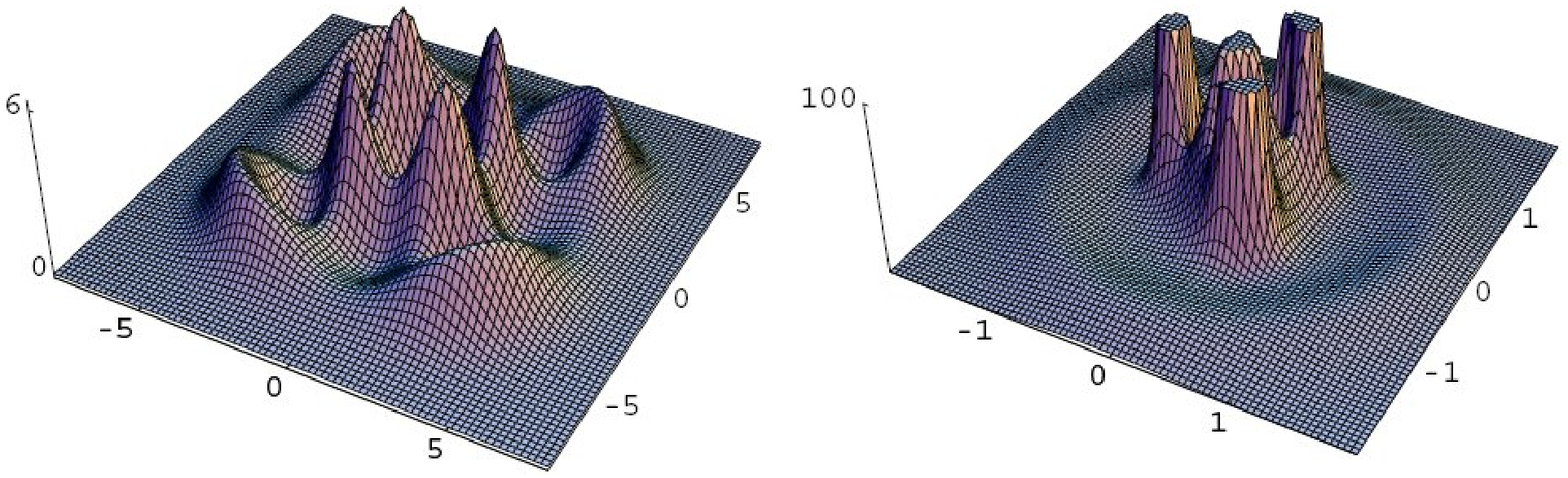}
}
\resizebox{8.5cm}{!}{
\includegraphics{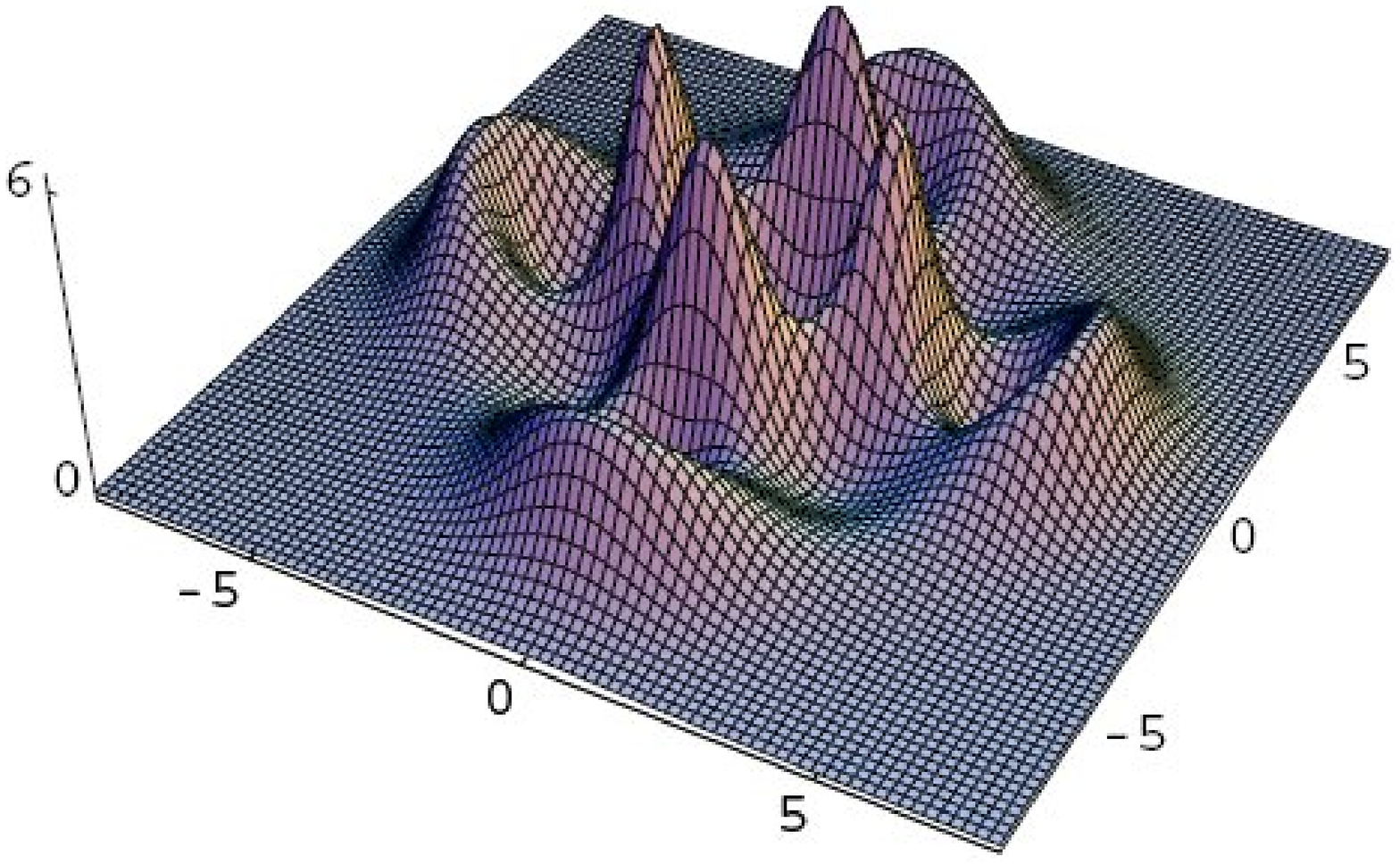}
}
\end{center} 
\caption{Plots of $\mathcal{E}$ at $t=-1000,0,1000$ for the solution of the 
example in section~\ref{subsec:general} with the data 
$p = \rmi$, $f = -\rmi \, \omega/40$, $h = \omega^4/10$ and $g = 0$. 
At $t=0$ we have cut off the lumps beyond a certain height. 
\label{fig:3Jordan_c} }
\end{figure}

\noindent
\textbf{Example.} Let 
\be
    P = \left( \begin{array}{cccccc} 
         p & 1 &   &        &        &         \\ 
         0 & p & 1 &        &        &         \\
           &   & p &        &        &         \\
           &   &   & p^\ast & 1      &         \\
           &   &   &        & p^\ast & 1       \\
           &   &   &        & 0      & p^\ast  
       \end{array} \right) , \; 
    T = \left( \begin{array}{cccccc}  
      &   &   &    &    & -1    \\ 
      &   &   &    & -1 &       \\ 
      &   &   & -1 &    &       \\
      &   & 1 &    &    &       \\
      & 1 &   &    &    &       \\
    1 &   &   &    &    & 
               \end{array} \right) , \; 
   V = \left( \begin{array}{cc}
              0 & 0 \\
              0 & 0 \\
              1 & 0 \\
              0 & 0 \\
              0 & 0 \\
              0 & 1 
              \end{array} \right) \nonumber \\
\ee
Then $\mathcal{X}'$ must have the form
\be
    \mathcal{X}' = \left( \begin{array}{cccccc} 
         f & \tilde{h} & \tilde{g} &        &                &                \\ 
         0 &  f        & \tilde{h} &        &                &                \\
           &           &   f       &        &                &                \\
           &           &           & f^\ast & \tilde{h}^\ast & \tilde{g}^\ast \\
           &           &           &        & f^\ast         & \tilde{h}^\ast \\
           &           &           &        &                & f^\ast  
       \end{array} \right) \, , 
\ee  
and (\ref{XY-lin-hier}) becomes
\be
 &&   f_{x_n} = p^{n-1} f_{x_1} \, , \qquad 
    \tilde{h}_{x_n} = p^{n-1} \tilde{h}_{x_1} + (n-1) p^{n-2} f_{x_1} \, , 
                                \cr
 &&  \tilde{g}_{x_n} = p^{n-1} \tilde{g}_{x_1} + (n-1) p^{n-1} \tilde{h}_{x_1} 
                + \frac{1}{2}(n-1)(n-2) p^{n-3} f_{x_1} \; .
\ee
Writing 
\be
    \tilde{h} = h + \frac{\pa f}{\pa p} \, , \qquad 
    \tilde{g} = g + \frac{\pa h}{\pa p}
                  + \frac{1}{2} \frac{\pa^2 f}{\pa p^2}  \, , 
\ee
with functions $f,g,h$, it follows that these equations are satisfied if 
the latter are arbitrary holomorphic functions of 
$\omega = \sum_{n \geq 1} p^n x_n$. Furthermore, we find that
\be
    K = \left( \begin{array}{cc}  0 & K_{12} \\ -K_{12}^\ast & 0  
        \end{array} \right) 
        \quad \mbox{with} \quad
    K_{12} = \left( \begin{array}{ccc}
             \rmi/\beta^3  & 3/\beta^4 & - 6 \rmi/\beta^5  \\
             1/\beta^2 & -2 \rmi/\beta^3  & -3/\beta^4 \\
            -\rmi/\beta & -1/\beta^2 & \rmi/\beta^3  
         \end{array} \right) 
\ee
and $\beta = 2 \Im(p)$, solves $[P,K] = Q$ with $Q=-V V^\dagger T$. 
The resulting class of solutions is regular since
\be
     \det(\mathcal{X}' - K) 
 &=& \beta^{-18} (1 + \beta^6 |f|^2)^3 
     + 2 \beta^{-12} (1 + \beta^6 |f|^2) | 3 \rmi \, f - \beta \tilde{h}|^2 
                    \nonumber \\
 && + \beta^{-6} | 2 f^2 + (f+ \rmi \, \beta \tilde{h})^2 + \beta^2 f \tilde{g} |^2 
     + \beta^{-12} | 6 f + \beta (4 \rmi \, \tilde{h} - \beta \tilde{g})|^2 \; . 
\ee
Now we have three arbitrary holomorphic functions at our disposal, so 
this class exhibits quite a variety of different structures. 
Figures \ref{fig:3Jordan_b} and \ref{fig:3Jordan_c} show some examples. 
If $h=g=0$, the typical behaviour is similar to the one shown in figure~\ref{fig:2Jordan_180degr_rot}. 
\hfill $\square$
\vspace{.1cm}

Comparing the data that determine the class of solutions in the example 
in section~\ref{subsec:anom_lumps}, based on a conjugate pair of 
$2\times 2$ Jordan normal form matrices $P_I$, with those of the last 
example, which is based on a conjugate pair of $3\times 3$ Jordan normal 
form matrices, there is an obvious generalization to the case of 
conjugate pairs of larger Jordan normal form matrices $P_I$. We note 
in particular that proposition~\ref{prop:K_IJ} can be generalized. 
Since the solutions turned out to be automatically \emph{regular} 
in the $2\times 2$ and $3\times 3$ case, it may well be that this holds 
in general. But a proof of this conjecture is out of reach so far.

\subsection{Superposing solutions}
\label{subsec:super}
The data that determine solutions of the $su(m)$ pdCM hierarchy 
on the basis of proposition~\ref{prop:su(m)-sols} are given by a set 
of matrices $(P,\mathcal{X}',T,V)$. 
Let two such sets be given, $(P_i,\mathcal{X}_i',T_i,V_i)$, $i=1,2$, 
with associated matrices $Q_i = -V_i V_i^\dagger T_i$, $K_i$ 
(as solutions of (\ref{Q=[P,K]})), and $\Phi_i$ given by (\ref{Phi-K-formula2}).
$P_i,\mathcal{X}_i,T_i$ are $N_i \times N_i$ matrices and $V_i$ is an 
$N_i \times m$ matrix. 
We can combine them into the larger matrices 
\be
 && P = \left(\begin{array}{cc} P_1 & 0 \\ 0 & P_2 \end{array}\right) \, , \qquad
     \mathcal{X}' = \left(\begin{array}{cc} \mathcal{X}'_1 & 0 \\ 
                                            0 & \mathcal{X}'_2
                          \end{array}\right) \, ,  \cr 
 &&  T = \left(\begin{array}{cc} T_1 & 0 \\ 0 & T_2 \end{array}\right) \, , 
         \qquad \; 
   V = \left(\begin{array}{c} V_1 \\V_2 \end{array}\right) \; .
\ee
Obviously, $(P,\mathcal{X}',T,V)$ again satisfies (\ref{X'-lin-hier}), 
(\ref{commPX'}), (\ref{X'dagg}), (\ref{P-su-cond}) and $T^\dagger=-T$. 
(\ref{Q_UV_su}) becomes
\be
    Q = \left(\begin{array}{cc} Q_1 & Q_{12}\\Q_{21} & Q_2 \end{array}\right) 
    \, ,\quad
    Q_{12} = -V_1 V_2^\dagger T_2 \, , \quad
    Q_{21} = -V_2 V_1^\dagger T_1 \, ,
\ee
and (\ref{Q=[P,K]}) with
\be
    K = \left( \begin{array}{cc}  K_1 & K_{12} \\ K_{21} & K_2  
        \end{array} \right)    \label{K-super}
\ee
yields the equations
\be
    P_1 K_{12} - K_{12} P_2 = Q_{12} \, , \qquad 
    P_2 K_{21} - K_{21} P_1 = Q_{21} \; .  \label{K12-eqs}
\ee
The off-diagonal blocks of $K$ are a source of complexity and non-triviality of
the resulting superposition. (\ref{K-su-cond}) then determines $K_{21}$ in 
terms of $K_{12}$ (or vice versa), 
\be
   K_{21} = T_2^{-1} K_{12}^\dagger T_1 \; .  \label{K21<->K12}
\ee
As a consequence, the second of equations (\ref{K12-eqs}) follows 
from the first. 
If we find a solution\footnote{Choosing $V_1$ and $V_2$ such that 
$V_1 V_2^\dagger =0$, we have $Q_{12}=0$ and (\ref{K12-eqs}) is solved by 
$K_{12}=0$. It follows that $\varphi$ is simply the sum of the solutions 
$\varphi_1$ and $\varphi_2$. But $V_1 V_2^\dagger =0$ also implies that 
$\varphi_1 \varphi_2 =0$, hence both constituent solutions $\varphi_1, \varphi_2$ 
must be degenerate, i.e. cannot have full rank. }
$K_{12}$ of the remaining equation, then we obtain 
\be
    \Phi = \left(\begin{array}{cc} \Phi_1  & \Phi_1 K_{12} \Phi_2 \\
            \Phi_2 K_{21} \Phi_1 & \Phi_2  \end{array}\right)
            \left(\begin{array}{cc} A_1^{-1} & 0 \\ 0 & A_2^{-1} 
                  \end{array}\right) \, , 
\ee
where
\be
    A_1 = \mathcal{I}_{N_1} - K_{12} \Phi_2 K_{21} \Phi_1  \, , \qquad
    A_2 = \mathcal{I}_{N_2} - K_{21} \Phi_1 K_{12} \Phi_2  \, , 
           \label{A1A2}
\ee
and $\varphi$ given by (\ref{varphi-VTPhiV}) solves the $su(m)$ 
pdCM hierarchy, provided that the inverses of $A_1$ and $A_2$ exist. 
If the two matrices $\Phi_i$ are regular (and thus also 
the corresponding solutions $\varphi_i$), then $\Phi$ and thus also $\varphi$ 
is regular if and only if $\det(A_1) \det(A_2) \neq 0$ (for all values 
of $x_1,x_2,\ldots$). 
Since $\det(A_1)=\det(A_2)$ by an application of Sylvester's determinant 
theorem, this reduces to the condition
\be
     \det(A_1) \neq 0 \; .  \label{super_reg_cond}
\ee
We note also that $\det(A_1)$ is real since
\be
      \det(A_1)^\ast 
  &=& \det(A_1^\dagger)
   = \det(\mathcal{I}_{N_1} - K_{21}^\dagger \Phi_2^\dagger 
      K_{12}^\dagger \Phi_1^\dagger)         \nonumber \\
  &=& \det(\mathcal{I}_{N_1} - T_1 K_{12} \Phi_2 K_{21} \Phi_1 T_1^{-1})
   = \det(A_1) \; .
\ee 
\vspace{.1cm}

\noindent
\textbf{Example.} We choose
\be
    P_1 &=& \left( \begin{array}{cccc} 
       p_1 & 1   &        &        \\ 
         0 & p_1 &        &        \\
           &     &p_1^\ast&  1     \\
           &     &  0     &p_1^\ast\\
       \end{array} \right) \, , \;
    \mathcal{X}'_1 = \left( \begin{array}{cccc} 
       f_1 & h_1 + \pa f_1/\pa p_1 & & \\ 
        0  & f_1 &        &            \\
           &     &f_1^\ast&h_1^\ast+(\pa f_1/\pa p_1)^\ast  \\
           &     &   0    &f_1^\ast    \\
       \end{array} \right) \, ,  \nonumber \\ 
   T_1 &=& \left( \begin{array}{cccc}  
    &   &    & -1 \\ 
    &   & -1 &    \\
    & 1 &    &    \\
  1 &   &    &    \\
               \end{array} \right) \, , \; 
   V_1 = \left( \begin{array}{cc}
              0 & 0 \\
              1 & 0 \\
              0 & 0 \\
              0 & 1 \\
              \end{array} \right) \, , 
\ee
where $f_1, h_1$ are arbitrary holomorphic functions of $\omega_1$ 
(with $\omega_1$ defined in (\ref{omega_j})), and 
\be
    P_2 = \left( \begin{array}{cc} 
            p_2 & 0       \\
            0   & p_2^\ast 
       \end{array} \right) \, , \; 
    \mathcal{X}'_2 = \left( \begin{array}{cc} 
            f_2 &  0  \\
            0   & f_2^\ast  
       \end{array} \right) \, , \; 
    T_2 = \left( \begin{array}{cc}  
           0 & -1 \\
           1 & 0 
               \end{array} \right) \, , \; 
   V_2 = \left( \begin{array}{cc}
              1 & 0 \\
              0 & 1 
              \end{array} \right) \, , \quad
\ee
with an arbitrary holomorphic function $f_2$ of $\omega_2$. 
Thus we superpose data corresponding to a regular solution of the 
kind treated in the example in section~\ref{subsec:anom_lumps} 
and data corresponding to a regular solution as given in example~1 
of section~\ref{subsec:class1}. In the following we assume that
\be
    p_1 \neq p_2  \, , \qquad  p_1 \neq p_2^\ast \; .   \label{p1/=p2}
\ee
Together with the conditions $p_i \neq p_i^\ast$, $i=1,2$, which the 
data of the components have to satisfy, this means that the constants 
$p_i$ and their complex conjugates are pairwise different. 
The second condition in (\ref{p1/=p2}) is in fact needed for the 
matrix $K$ to exist.
$K$ has the form (\ref{K-super}), where $K_1$ is given by the 
$4 \times 4$ matrix $K$ in (\ref{K_2intlumps}) with the pair 
$(p_1,p_1^\ast)$. Furthermore, 
\be
    K_2 = \left( \begin{array}{cc} 0 & \frac{1}{p_2-p_2^\ast} \\
                                \frac{1}{p_2-p_2^\ast} & 0
                    \end{array} \right) , \; 
    K_{21} = \left( \begin{array}{cccc}
              0 & 0 & -\frac{1}{p_1^\ast - p_2} & \frac{1}{(p_1^\ast - p_2)^2} \\
              \frac{1}{p_1 - p_2^\ast} & -\frac{1}{(p_1 - p_2^\ast)^2} & 0 & 0
                    \end{array} \right) , 
\ee
and $K_{12}$ is then determined by (\ref{K21<->K12}). 
With some efforts the expression for $\det(A_1)$ can be 
brought into the form
\be 
     \det(A_1) 
  &=& |a|^{-8} (1 + \beta_2^2 |f_2|^2)^{-1} 
     \left( |w|^2 + (1+ \beta_1^4 |f_1|^2)^2 \right)^{-2} 
     \Big[ \Big( |a|^4 [ \, |w|^2 + (1+ \beta_1^4 |f_1|^2)^2 \, ] 
              \nonumber \\
  && - \beta_1 \beta_2 \, [ \, |a|^2 (1 + \beta_1^4 |f_1|^2 ) 
     + |b|^2 + |a w + \rmi b^\ast \beta_1^2 f_1|^2 \, ] \Big)^2 
               \nonumber \\
  && + \Big| (a^\ast)^4 ( |w|^2 + (1 + \beta_1^4 |f_1|^2)^2 ) \beta_2 f_2 
              \nonumber  \\
  &&   + \beta_1 \beta_2 \Big( b^2 w + (a^\ast)^2 \beta_1^4 f_1^2 w^\ast 
     - 2 \rmi a^\ast b \beta_1^2 f_1 ( 1 + \beta_1^4 |f_1|^2) \Big) \Big|^2 
     \, \Big] \, , 
\ee
where
\be
    \beta_i = 2 \, \Im(p_i) \, , \quad
    a = p_1 - p_2^\ast \, , \quad
    b = p_1 - p_2 \, , \quad 
    w = \beta_1^3 h_1 - 2 \rmi \, \beta_1^2 f_1 \; . 
\ee
The regularity condition (\ref{super_reg_cond}) turns out to be 
automatically satisfied. This is seen as follows. First we note that 
\be
     |a|^4 = |b|^4 + \beta_1 \beta_2 \, (|a|^2+|b|^2) 
           > \beta_1 \beta_2 \, (|a|^2+|b|^2) \, , 
\ee
as a consequence of the first of the inequalities (\ref{p1/=p2}), and thus  
\be
   |a|^4 \, (1 + \beta_1^4 |f_1|^2)^2 
    > \beta_1 \beta_2 \, (|a|^2+|b|^2)(1+\beta_1^4 |f_1|^2) \; .  
\ee
Using $|a|^2 > \beta_1 \beta_2$, this leads to 
\be
      |a|^4 \, ( |w|^2 + (1+ \beta_1^4 |f_1|^2)^2 ) 
  &>& \beta_1 \beta_2 \, \left( |a|^2 (1+\beta_1^4 |f_1|^2)+|b|^2 
      + |a w|^2 + |\rmi \, b^\ast \beta_1^2 f_1|^2 \right) \nonumber \\
  &\geq& \beta_1 \beta_2 \, \left( |a|^2 (1+\beta_1^4 |f_1|^2)+|b|^2 
      + |a w + \rmi \, b^\ast \beta_1^2 f_1|^2 \right) \, ,  
\ee
which implies $\det(A_1) >0$.

\begin{figure}[t] 
\begin{center} 
\resizebox{16cm}{!}{
\includegraphics{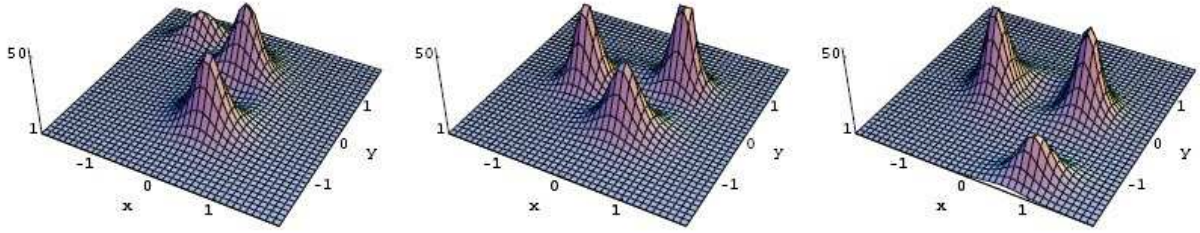}
}
\end{center} 
\caption{Plots of $\mathcal{E}$ at $t=-2,0,2$ for a superposition of a  
2-lump configuration, with ``anomalous scattering'', and a single lump 
(which is at the top of the left plot and at the bottom of the right plot), 
according to the example of section~\ref{subsec:super}.  
Here we chose $p_1 = \rmi$, $p_2=-3\rmi/8$, $f_1 = -\rmi \, \omega_1/32$, 
$f_2 = 24 \, \omega_2$, and $h_1 = -\omega^2/4$. 
\label{fig:2+1lumps} }
\end{figure}

Figure~\ref{fig:2+1lumps} shows plots of 
$\mathcal{E}$ at consecutive times, for a special choice of the data. 
\hfill $\square$

\vspace{.1cm}
In the last example the regularity of the superposition turned out to be 
a consequence of the ``regular data'' we started with. But this 
example also demonstrates that it is quite difficult in general 
to evaluate the regularity condition (\ref{super_reg_cond}). 
We note that also the cases treated in sections~\ref{subsec:regsols} and \ref{subsec:anom_lumps} may be regarded as special cases of 
``superpositions'' as formulated above. In particular, example~2 of 
section~\ref{subsec:regsols} provides us with another example where 
the superposition of regular data turned out to be regular again. 
It is unlikely that this is a special feature of our particular examples.
But in order to tackle a general proof, we probably need different methods.

\section{Conclusions}
\label{section:conclusions}
We summarize the relations between integrable systems and their hierarchies considered in this work in the following diagram. 
\vskip.5cm

\[
 \setlength{\arraycolsep}{2.cm}
 \begin{array}{cc}
   \Rnode{a}{\mathrm{mKP}_Q} & \Rnode{b}{\mathrm{Ward}_Q} \\[1.cm]
   \Rnode{c}{\mathrm{pKP}_Q} & \Rnode{d}{\mathrm{pdCM}_Q} \\[1.cm]
   \Rnode{e}{\mathrm{scalar} \; \mathrm{pKP}} & \Rnode{f}{su(m) \; \mathrm{pdCM}} 
 \end{array}
 \psset{nodesep=5pt,arrows=->,linestyle=dashed}
 \ncLine{a}{b}\Aput{\mbox{dispersionless limit}}
 \psset{nodesep=5pt,arrows=->,linestyle=solid}
 \ncLine{c}{d}\Aput{\mbox{dispersionless limit}}
 \ncLine{c}{e}\Bput{\mbox{rank}(Q)=1 } 
 \ncLine{c}{e}\Aput{\mbox{reality cond.} }
 \ncLine{d}{f}\Bput{\mbox{rank}(Q)=m }
 \ncLine{d}{f}\Aput{su(m) \; \mbox{cond.} }
 \psset{nodesep=5pt,arrows=<->,linestyle=dashed}
 \ncLine{a}{c}\Bput{\mbox{Miura transf.}}
 \psset{nodesep=5pt,arrows=<->,linestyle=solid}
 \ncLine{b}{d}\Aput{\mbox{pseudo-duality}}
\]

\noindent
Here $\mathrm{pKP}_Q$ and $\mathrm{pdCM}_Q$ stand, respectively, for the 
pKP and pdCM hierarchy in the matrix algebra $GL(M \times N,\mathbb{C})$ 
with product modified by a constant matrix $Q$ (see (\ref{Qproduct})). 
$\mathrm{pdCM}_Q$ is related by pseudo-duality (see (\ref{varphi-J-eq})) to 
the hierarchy $\mathrm{Ward}_Q$ of Ward's 
model with dependent variable in $GL(M \times N,\mathbb{C})$ (and product 
modified by $Q$). If $\mbox{rank}(Q)=1$, solutions of $\mathrm{pKP}_Q$ 
are mapped to solutions of the scalar pKP hierarchy, an additional  
condition ensures that the resulting solution is real. 
Analogously, if $\mbox{rank}(Q)=m$ and an $su(m)$ condition holds, 
solutions of the $\mathrm{pdCM}_Q$ hierarchy are mapped to solutions 
of the $su(m)$ pdCM hierarchy (which is pseudo-dual to the hierarchy 
associated with Ward's modified $SU(m)$ chiral model). 
Concerning the Miura transformation between the (matrix) pKP hierarchy 
and the modified KP (mKP) hierarchy, and its dispersionless limit 
(see the dashed arrows in the diagram), see \cite{DMH06func}.  
The relations provided by the dispersionless scaling limits in the diagram 
have actually been anticipated in \cite{DMH06func} (see the remark in 
section~4 therein). 

In the present work we demonstrated how the dispersionless scaling can 
be used to transfer a method of constructing exact solutions from the 
(matrix or ``noncommutative'') pKP hierarchy to the pdCM hierarchy. Indeed, 
proposition~\ref{prop:pdcm_sols} is an analog of theorem~4.1 in 
\cite{DMH07Burgers} (which we recalled as theorem~\ref{theorem:KPCH}). 
We showed that large classes of exact solutions of the 
pdCM hierarchy can be obtained with its help. In particular, 
we presented examples of various multiple lump configurations of the 
$su(m)$ pdCM. 
The general result formulated in proposition~\ref{prop:pdcm_sols} is 
a source of even more classes of exact solutions. 

Our method to generate exact solutions of the $su(m)$ pdCM hierarchy 
is based on quite simple formulae and quickly produces interesting 
solutions (like lumps with ``anomalous scattering''). But a more systematic 
treatment, in particular of multi-lump solutions, requires deeper methods 
(of matrix calculus), and further insights are needed as to 
how the a priori given plethora of parameters can efficiently be reduced. 
It would also be of interest to compare this method with an inverse 
scattering approach. 

Solutions of the $su(m)$ pdCM hierarchy can also be obtained from 
solutions of Ward's chiral model hierarchy by integrating 
(\ref{varphi-J-eq}) (or equivalently (\ref{varphi-J-eq2})). 
In any case, one should expect an analogous structure of localized solutions,  
and this indeed turns out to be the case in examples. A deviation 
in the corresponding plots is caused by the fact that the ``energy'' 
expression for the pdCM differs from the energy of the Ward model 
by a term that causes an asymmetry in the $x$-direction, 
see the remark in section~\ref{subsec:dncpKP_Ward} and also the 
appendix.\footnote{We note that $x$ and $y$ have to be exchanged for 
comparing our formulae with those in the literature on the Ward model.}
Our method to generate solutions of the pdCM hierarchy seems to be quite 
different from the methods that were used to construct solutions of 
Ward's model. In particular, in the latter model solutions with 
``anomalous scattering'' have been obtained by taking suitable limits 
of families of non-interacting lump solutions. In our approach, 
corresponding solutions of the pdCM hierarchy are directly given 
by matrix data involving Jordan blocks. Moreover, we have seen that 
even the simple multiple lump solutions of section~\ref{subsec:regsols} 
can exhibit an anomalous behaviour within some compact space region 
(see figure~\ref{fig:2lumps_int}), whereas asymptotically (i.e. compared 
at large enough negative and positive times) no deflection is observed. 
We noted that this has a KP-I counterpart \cite{LTG04} and also 
analogs in some other systems \cite{McWi+Zabu82,Gibb+Mant86}. 

The fact that the dispersionless scaling limit of matrix pKP (respectively  
mKP) is a simple reduction of a potential version of the (4-dimensional) 
self-dual Yang-Mills equation raises the question whether there is a 
(4-dimensional) integrable system that has the full self-dual Yang-Mills 
equation as a dispersionless limit and that admits a reduction to 
matrix KP (respectively mKP).

\appendix

\section*{Appendix: The ``anti-pdCM hierarchy''}
\setcounter{section}{1}
Writing (\ref{Ward-hier}) as $J \rmd(J^{-1} \bd J)J^{-1}=0$, using the 
Leibniz rule and (\ref{bicomplex}), we obtain the equivalent form 
\be
     \bd( (\rmd J) J^{-1}) = 0 
\ee
of the hierarchy associated with Ward's chiral model, where $\rmd$ and $\bd$ 
exchanged their roles. This is integrated 
by introducing a potential $\tv$ such that
\be
    (\rmd J) J^{-1} = \bd \tv \; .   \label{tvarphi}
\ee
Rewriting the last equation in the form $\rmd J = (\bd \tv) J$, 
we obtain the integrability condition
\be
    \rmd \bd \tv = \bd \tv \wedge \bd \tv \; .  \label{anti-pdCM-hier}
\ee
In components, this becomes
\be
    \tv_{x_n x_{m+1}} - \tv_{x_m x_{n+1}} = [\tv_{x_{n+1}} , \tv_{x_{m+1}}] 
    \qquad \quad  m,n=1,2,\ldots \, ,
\ee
which we refer to as the \emph{anti-pdCM hierarchy}. 
For $m=1$ and $n=2$ we have
\be
    \tv_{x_1 x_3} - \tv_{x_2 x_2} = [\tv_{x_2},\tv_{x_3}] \; .  
\ee
In terms of the coordinates given by (\ref{cord_xyt}), it takes the form
\be
    \tv_{t t} -\tv_{x x}-\tv_{y y} + [\tv_t + \tv_x,\tv_y] = 0 \; .
\ee
Since this equation is obtained from (\ref{1st-eq-xyt}) by $x \mapsto -x$, 
so are its Lagrangian $\tilde{\mathcal{L}}$ and energy-momentum tensor 
$\tilde{T}^\mu{}_\nu$ from those in section~\ref{subsec:dncpKP_prop}. 
For $\tv$ in $su(m)$, 
\be
    \tilde{\mathcal{E}} 
  = \tilde{T}^0{}_0 + \tilde{T}^0{}_1 
  = - \frac{1}{2} \tr\Big( (\tv_t +\tv_x)^2+\tv_y{}^2 \Big) 
\ee
is then a \emph{non-negative} conserved density. 

Associated with any solution $J$ of Ward's chiral model hierarchy via 
(\ref{varphi-J-eq2}) and (\ref{tvarphi}), there are solutions $\varphi$ and 
$\tv$ of the pdCM hierarchy and the anti-pdCM hierarchy, respectively. 
Using $J_t J^{-1} - J_x J^{-1} = \tv_y$ and $J_y J^{-1} = \tv_t + \tv_x$, 
which follow from (\ref{tvarphi}), we find that 
\be
    \tilde{\mathcal{E}} 
  = \mathcal{E}_{\mathrm{Ward}} + \tr\Big( J^{-1} J_t \,J^{-1} J_x \Big) \; .
\ee
Combining this with (\ref{E-E_Ward}), leads to
\be
   \mathcal{E}_{\mathrm{Ward}} = \frac{1}{2}(\mathcal{E}+\tilde{\mathcal{E}}) \; .
\ee

The next result is an analogue of corollary~\ref{cor:Phi-K-formula} and 
leads to a class of solutions of the anti-pdCM hierarchy.

\begin{propA}
Let $(P,K,\mathcal{X}')$ be data that determine via 
corollary~\ref{cor:Phi-K-formula} a solution 
$\Phi$ (given by (\ref{Phi-K-formula2})) of the pdCM hierarchy with $Q=[P,K]$. 
If $P$ is invertible (as in all our examples in section~\ref{section:pdc-sols}), 
then $\Phi$ also solves  
\be
    \rmd \bd \Phi = \bd \Phi \wedge \tilde{Q} \bd \Phi  \label{tQ-anti-pdCM-hier}
\ee
with $\tilde{Q} = P^{-1} Q P^{-1}$.
\end{propA}
\textbf{Proof:} As a consequence of (\ref{X'-lin-hier}), 
$\mathcal{X} := \mathcal{X}' -K$ solves 
\bez
    \rmd \mathcal{X} = \bd \mathcal{X} \, P^{-1} \, , 
\eez
and consequently $\Phi = \mathcal{X}^{-1}$ satisfies
\bez
    \rmd \Phi = (\bd \Phi) \, W^{-1} \, , \qquad 
    W = \mathcal{X} P \mathcal{X}^{-1} \; .
\eez
Now we note that (\ref{commPX'}) and (\ref{Q=[P,K]}) imply
$[P^{-1},\mathcal{X}] = [K,P^{-1}] = P^{-1} Q P^{-1}$. Hence
\bez
    W^{-1} = P^{-1} (\mathcal{X} + Q P^{-1} ) \, \mathcal{X}^{-1} 
           = P^{-1} + P^{-1} Q P^{-1} \Phi \, , 
\eez
and we obtain
\bez
    \rmd \bd \Phi = - \bd \rmd \Phi  
  = (\bd \Phi) \wedge P^{-1} Q P^{-1} \bd \Phi \; .
\eez
\hfill $\square$

If moreover the assumptions of proposition~\ref{prop:su(m)-sols} are satisfied, 
then
\be
    P^{-1} Q P^{-1} = - P^{-1} V V^\dagger T P^{-1}
 = -P^{-1} V (P^{-1} V)^\dagger T  \, , 
\ee
and (\ref{tQ-anti-pdCM-hier}) implies that
\be
    \tv = - (P^{-1} V)^\dagger T \Phi P^{-1} V   \label{su-tv}
\ee
solves the anti-Hermitian anti-pdCM hierarchy (\ref{anti-pdCM-hier}). 
The data $(P,\mathcal{X}',T,V)$ therefore determine a solution 
(\ref{varphi-VTPhiV}) of the anti-Hermitian pdCM hierarchy and also a 
solution (\ref{su-tv}) of the anti-Hermitian anti-pdCM hierarchy.

Although elaboration of examples suggests 
that the pair $(\varphi, \tv)$ determined by the data $(P,\mathcal{X}',T,V)$ 
indeed belongs to the same solution $J$ of Ward's chiral model hierarchy 
(via (\ref{varphi-J-eq2}) and (\ref{tvarphi})), 
we were not able so far to prove this.

\end{document}